\documentclass[journal=tches,final]{iacrtrans}

\usepackage{lipsum} 
\usepackage{graphicx}
\usepackage{float}
\usepackage{enumitem}
\usepackage{amssymb}
\usepackage{amsmath}
\usepackage[linesnumbered,ruled]{algorithm2e}
\usepackage{array}
\usepackage{multirow}
\usepackage{makecell}
\SetKwInput{KwInput}{Input}                
\SetKwInput{KwOutput}{Output}              

\graphicspath{ {./images/} }
\author{Zhehong Wang, Dennis Sylvester, Hun-Seok Kim, \and David Blaauw}
\institute{
  University of Michigan, Ann Arbor, USA, \email{{zhehongw, dmcs, hunseok, blaauw}@umich.edu}
}

\title{Hardware Acceleration for Third-Generation FHE and PSI Based on It}

\begin{document}

\maketitle


\begin{abstract}
With the expansion of cloud services, serious concerns about the privacy of users' data arise due to the exposure of the unencrypted data to the server during computation. Various security primitives are under investigation to preserve privacy while evaluating private data, including Fully Homomorphic Encryption (FHE), Private Set Intersection (PSI), and others. However, the prohibitive processing time of these primitives hinders their practical applications. This work proposes and implements an architecture for accelerating third-generation FHE with Amazon Web Services (AWS) cloud FPGAs, marking the first hardware acceleration solution for third-generation FHE. We also introduce a novel unbalanced PSI protocol based on third-generation FHE, optimized for the proposed hardware architecture. Several algorithm-architecture co-optimization techniques are introduced to allow the communication and computation costs to be independent of the Sender's set size. The measurement results show that the proposed accelerator achieves $>21\times$ performance improvement compared to a software implementation for various crucial subroutines of third-generation FHE and the proposed PSI.
\end{abstract}

\section{Introduction}
\label{sec:intro}

The past decades have witnessed the increasingly wide deployment of data centers. With their excellent storage capacity and computational performance, they have assumed a fundamental role in almost every aspect of human life. Nowadays, it is common to outsource data or computation extensive workloads to large data centers. However, for servers to work on the data from a client, the encrypted client data must be decrypted first, which reveals the private data to a potentially untrusted cloud server. Thus, as a growing number of services are moving online, especially after the COVID-19 pandemic, privacy-preserving computation, which allows secure operation on encrypted client data, is expected to play a pivotal part of this industry in the future. In this work, two privacy-preserving schemes and their hardware acceleration are discussed.

\subsection{Fully Homomorphic Encryption}
\label{sec:introsub1}
Fully Homomorphic Encryption (FHE) \cite{FHE1, FHE2, FHE3, FHE4, FHE5, FHE6, FHE7, FHE8, FHE9, FHE10, FHE11}, which permits secure computation on encrypted data without decrypting it, has been given extensive research in recent years to enable privacy-preserving computation. To be precise, for a given function $f(x)$, a homomorphic encryption scheme satisfies $f(Enc(m))= Enc(f(m))$. If it is homomorphic to any function, it is characterized as fully homomorphic encryption.

The very first FHE scheme was not devised until 2009, when Gentry proposed a general FHE framework \cite{FHE1, FHE2}. It has been proven that from a Boolean circuit model perspective of computation, if an encryption scheme is homomorphic to its own decryption function followed by a universal logic gate, then it is homomorphic to any function. The operation that fulfills this property by refreshing the noise level of the ciphertext after each operation, thus transforming a Leveled HE (LHE) into an FHE, is called bootstrapping or recryption. Based on this idea, Gentry also presented a concrete construction that takes around 30 minutes per bootstrapping \cite{FHE3}. 

Following Gentry's blueprint, various schemes have been proposed for better efficiency. Among them, the most well-known are BGV \cite{FHE4}, BFV \cite{FHE5, FHE6}, and CKKS \cite{FHE7}. These second-generation schemes differ from Gentry’s approach in relying on the Ring Learning with Errors (RLWE) problem for its better-studied hardness analysis and efficiency obtained via SIMD-styled computation \cite{FHE8}. These schemes focus on evaluating polynomials homomorphically. Several open-source implementations \cite{PALISADE, HElib, SEAL} can potentially reduce the recryption time to minutes depending on the security parameters.

After the proposal of GSW \cite{FHE9} in 2013, FHEW \cite{FHE10} and TFHE \cite{FHE11} were published as third-generation approaches. The third-generation schemes focus on efficient homomorphic boolean logic evaluation. Although performance-wise the third generation may not be superior to earlier schemes (since the amortized cost of the SIMD-styled second generation is estimated to be within the same order of magnitude as that of the third generation), the third generation is well accepted for its simplicity and flexibility in terms of both concept and implementation. Typically, to achieve a 128-bit security level, the degree of the polynomial is less than 2048, which is much shorter than the \textasciitilde10,000 in the second generation. Further, the ciphertext modulus is less than 64 bits, compared to \textasciitilde200 bits in the second generation. The reported recryption time of the third generation is generally around 0.1s to 1s \cite{FHEWlike}.

Although several improvements have been presented to increase the efficiency of FHE, secure computation on encrypted data is still many orders of magnitudes slower compared to direct computation on plaintext, due to the prohibitive requirement of compute power that challenges current computing systems, which hinders practical application of FHE. For example, the time of bootstrapping one homomorphic NAND gate is sub-seconds  for third-generation schemes \cite{FHE10, FHE11}, compared to picoseconds of a CMOS NAND.This still renders FHE largely impractical. Thus, various hardware solutions have been proposed in recent years \cite{FHEHW1, FHEHW2, FHEHW3, FHEHW4, FHEHW5, FHEHW6, FHEHW7, FHEHW8, FHEHW9, FHEHW10, FHEHW11}. \cite{FHEHW1, FHEHW3} focused on encryption/decryption of RLWE in a post-quantum scenario, which is less computation demanding due to the smaller size of the polynomial. Reference \cite{FHEHW2} presented a crypto-engine for the encryption/decryption of RLWE for homomorphic encryption, which is less heavy lifting compared to homomorphic evaluation. The authors in \cite{FHEHW10} explored acceleration for large number multiplication, while \cite{FHEHW8, FHEHW9} discussed approaches to accelerate long polynomial multiplications in homomorphic encryption. Other works \cite{FHEHW5, FHEHW6, FHEHW7, FHEHW11} implemented accelerators for LHE schemes based on the BFV scheme, with limited computation depth and security levels. Finally, an architecture for the CKKS scheme was proposed in \cite{FHEHW4}. To date, there has been no hardware acceleration published for third-generation FHE schemes.

\subsection{Private Set Intersection}
\label{sec:introsub2}
PSI is another security primitive that preserves privacy of operation. It allows two parties (Sender and Receiver) to exchange the intersection of their private sets without leaking any excess information other than the intersection set. Thus, its applications include private human genome testing, contact list discovery of social media apps, and conversion rate measuring of online advertisement. Recently, Microsoft introduced Password Monitor in the latest release of the Edge web browser, which compares a user’s private passwords saved to Edge with a known database of leaked passwords to figure out whether there is a leak in the user’s passwords. With the underlying PSI protocol, the server that facilitates the comparison learns nothing about the user’s passwords.

The PSI problem has been explored extensively, seeking efficient protocols \cite{PSI1, PSI2, PSI3, PSI4, PSI5}. However, in an unbalanced scenario where one set is significantly smaller than the other, these protocols perform linearly on the size of the large set. In recent years, unbalanced PSI protocols based on second-generation FHE \cite{PSI6, PSI7} were proposed that provide significant communication overhead reduction compared to previous approaches but maintain comparable performance. However, they still suffer from the large encryption parameters of second-generation FHE. While  third-generation FHE can perform boolean logic more efficiently, it is a natural candidate for performing the comparison in PSI. However, this has not been explored before.

\subsection{Our Contributions}
\label{sec:introsub3}
This paper presents the following contributions: 
\begin{itemize}
	\item We present the first accelerator architecture for third-generation FHE, targeting the $RLWE\bigotimes RGSW$ operation (defined in the following section), which is a fundamental function of both second-generation and third-generation FHE. By exploiting the asymmetric nature of the operation, the architecture is capable of maintaining high throughput with less resource usage while addressing different parameter sets. An extensive analysis of the architecture is included.
	\item We propose a novel unbalanced PSI protocol that is based on third-generation FHE and is demonstrated with the proposed hardware. The proposed PSI protocol makes the computation cost independent of the Sender’s set size. The core block of the PSI that facilitates the cross comparison of the PSI in \cite{PSI6} is replaced with a homomorphic lookup table (LUT) implemented with third-generation FHE. Unlike the multiplication used in \cite{PSI6}, which returns a nonzero value when the cross comparison misses and potentially leaks the content of the Sender’s set, the LUT only returns one bit indicating whether an element is inside the Sender’s set; and thus, avoiding sending any excess information about the Sender’s set. Therefore, the noise flooding process adopted in \cite{PSI6} is not necessary. We introduce several additional algorithm-architecture co-optimizations to reduce the computation and communication costs, rendering a practical application of the proposed PSI protocol.
	\item A prototype of the proposed architecture is implemented with AWS cloud FPGA service. We develop all necessary high-level functions in C++ and benchmark the implemented architecture with different parameter sets. We make the SystemVerilog HDL code of the proposed accelerator and supporting software code publicly available at \cite{MYREPO}.
	\item We quantify and analyze the performance of the proposed hardware acceleator and PSI protocol. The measurements show over $21\times$ performance improvement compared to a software implementation for various subroutines of the third-generation FHE and the proposed PSI.
\end{itemize}

\section{Preliminaries}
\label{sec:background}

\subsection{Notation}
\label{sec:backgroundsub1}
Throughout the paper, boldface lower-case letters $\boldsymbol{a,b,c,} \dots$ are used to denote vectors or polynomials depending on the context, and boldface upper-case letters $\boldsymbol{A,B,C,} \dots$ are used for matrices. The set of integers is denoted by $\mathbb{Z}$, and the quotient ring of integers modulo $q$ is denoted by $\mathbb{Z}_q$. The polynomial ring is denoted by $R=\mathbb{Z}[X]/(X^N+1)$, where N is a power of two. And $R_Q=R/QR$ represents the residue ring of $R$ modulo an integer $Q$. “$\times$” denotes the scalar multiplication with either another scalar or a vector/polynomial. “$\cdot$” denotes the vector inner product or polynomial product depending on the context, while “$\bigodot$” denotes the outer product or element wise product of a polynomial. Lastly, “$RLWE \bigotimes RGSW$” represents the product of an RLWE ciphertext and an RGSW ciphertext, which will be detailed in the next section. 

\subsection{Lattice-based Cryptography: LWE, RLWE and RGSW}
\label{sec:backgroundsub2}
Almost all FHE schemes published so far are built upon the LWE and/or RLWE problem, which can be reduced to a lattice problem that is proven to be quantum safe within polynomial time \cite{LATTICE, IDEALATTICE}.

\subsubsection{Learning with Errors Encryption}
In practice, given a plaintext modulus $t$ and a ciphertext modulus $q$, an LWE encryption of a plaintext $m\mod t$ with secret vector $\boldsymbol{s}$ is defined as:
\begin{equation}
	LWE_{\boldsymbol{s}}^{q/t}(m)=[\boldsymbol{a},b=\boldsymbol{a}\cdot \boldsymbol{s}+e+m\times q/t\mod q]
	\label{eq:LWE}
\end{equation}
with the vector $\boldsymbol{a}$, of dimension $n$, sampled uniformly from the integer vector space $\mathbb{Z}_q^n$ and error $e$ sampled from an error distribution $\chi$ \cite{LATTICE}. As long as $|e|<q/(2t)$, the plaintext can be successfully recovered by $m=\lfloor (b-\boldsymbol{a\cdot s})\times t/q\rceil $, which rounds off the noise.

\subsubsection{Ring Learning with Errors Encryption}
\label{sec:backgroundsub2sub2}
However, additive homomorphism is not enough to construct the bootstrap function. RLWE that is potentially multiplicative homomorphic is also incorporated in third-generation FHE. Similar to the definition of LWE, given a plaintext modulus $T$ and a ciphertext modulus $Q$, an RLWE encryption of a plaintext polynomial $\boldsymbol{m} \mod T$ with secret polynomial $\boldsymbol{s}$ is defined as follows:
\begin{equation}
	RLWE_{\boldsymbol{z}}^{Q/T} (\boldsymbol{m})=[\boldsymbol{a},\boldsymbol{b}=\boldsymbol{a}\cdot\boldsymbol{z}+\boldsymbol{e}+\boldsymbol{m}\times Q/T  \mod Q]
	\label{eq:RLWE}
\end{equation}
with the polynomial $\boldsymbol{a}$ sampled uniformly from the ring $R_Q$, and $\boldsymbol{e}$, a noise polynomial, sampled from an error distribution $\chi$ \cite{IDEALATTICE}. As long as $|\boldsymbol{e}|_\infty<Q/(2T)$, the plaintext can be successfully recovered by $m=\lfloor (\boldsymbol{b}-\boldsymbol{a\cdot z})\times T/Q\rceil$, which rounds off the noise. In some contexts, the scale $Q/T$ is omitted for clarity.

Since RLWE is a special form of LWE, the coefficients of the polynomial $\boldsymbol{b}$ of an RLWE ciphertext $RLWE_{\boldsymbol{z}}(\boldsymbol{m})=[\boldsymbol{a,b}]$ can be converted into multiple separate LWE ciphertexts under the same secret key with some transformation of polynomial $\boldsymbol{a}$, which is detailed in Appendix \ref{sec:appRLWE2LWE}.

\subsubsection{NTT and INTT}
The polynomial multiplication of RLWE can be efficiently computed with NTT. NTT is an adaptation of the well-known FFT algorithm, which reduces the complexity of polynomial multiplication from $O(N^2)$ to $O(Nlog(N))$. However, to perform polynomial multiplication modulo $(X^N+1)$, negacyclic/anticyclic convolution is adopted  \cite{NTTTRICK}. The optimized NTT/INTT algorithms summarized in \cite{NTTTRICK} are adopted and implemented on hardware in this work. The algorithms are given in Appendix \ref{sec:appNTT} and \ref{sec:appINTT}.

\subsubsection{Ring GSW Encryption}
\label{sec:backgroundsub2sub4}
Lastly, Ring-GSW (RGSW) encryption is widely adopted in third-generation FHE to facilitate homomorphic polynomial multiplication \cite{FHE10, FHE11}. It is defined as a matrix of RLWEs (in some literature, the two columns are concatenated as a one-dimensional vector): 
\begin{equation}
	RGSW_{\boldsymbol{z}} (\boldsymbol{m})=[RLWE_{\boldsymbol{z}}^{'}(-\boldsymbol{z\cdot m}),RLWE_{\boldsymbol{z}}^{'} (\boldsymbol{m})],
	\label{eq:RGSW}
\end{equation}
with $RLWE_{\boldsymbol{z}}^{'} (\boldsymbol{m})$ defined as a vector of RLWEs:
\begin{equation}
	RLWE_{\boldsymbol{z}}^{'} (\boldsymbol{m})=[RLWE_{\boldsymbol{z}}(\boldsymbol{m}),RLWE_{\boldsymbol{z}} (B_G\times\boldsymbol{m}),\cdots,RLWE_{\boldsymbol{z}} (B_G^{dc-1}\times\boldsymbol{m})],
	\label{eq:RLWEPrime}
\end{equation}
where $B_G$ is a predefined decomposition base and $dc=log_{B_G}Q$ denotes the length of vector $RLWE_{\boldsymbol{z}}^{'}(\boldsymbol{m})$. 

The multiplication of an $RLWE(\boldsymbol{m_1})=[\boldsymbol{a,b}]$ and an $RGSW(\boldsymbol{m_2})$ is defined in \autoref{eq:RLWEtimesRGSW}, with the two polynomials of the RLWE being decomposed by the base $B_G$ into two vectors of polynomials, $\boldsymbol{a^{'}}$ and $\boldsymbol{b^{'}}$, that satisfy $\boldsymbol{a}=\sum_{i=0}^{dc-1}\boldsymbol{a^{'}}[i]$ and $\boldsymbol{b}=\sum_{i=0}^{dc-1}\boldsymbol{b^{'}}[i]$. Further, the product of a polynomial $p$ and an RLWE ciphertext  $RLWE_{\boldsymbol{z}} (\boldsymbol{m})=[\boldsymbol{a,b}]$ is defined as $\boldsymbol{p}\cdot RLWE_{\boldsymbol{z}} (\boldsymbol{m})=[\boldsymbol{p\cdot a,p\cdot b}]$. The $\bigotimes$ operator is used extensively in the bootstrap process, and is the main focus of our hardware implementation. 

\begin{equation}
	\begin{split}
		RLWE(\boldsymbol{m_1})\bigotimes RGSW(\boldsymbol{m_2})&=\boldsymbol{a^{'}}\cdot RLWE^{'} (-\boldsymbol{z\cdot m_2})+\boldsymbol{b^{'}}\cdot RLWE^{'} (\boldsymbol{m_2}) \\
		&=RLWE(\boldsymbol{m_1\cdot m_2+e_1\cdot m_2})
	\end{split}	
	\label{eq:RLWEtimesRGSW}
\end{equation}

\begin{figure}
	
	\caption{Data Flow of Homomorphic Accumulation in FHEW.}
	\includegraphics[scale=0.46]{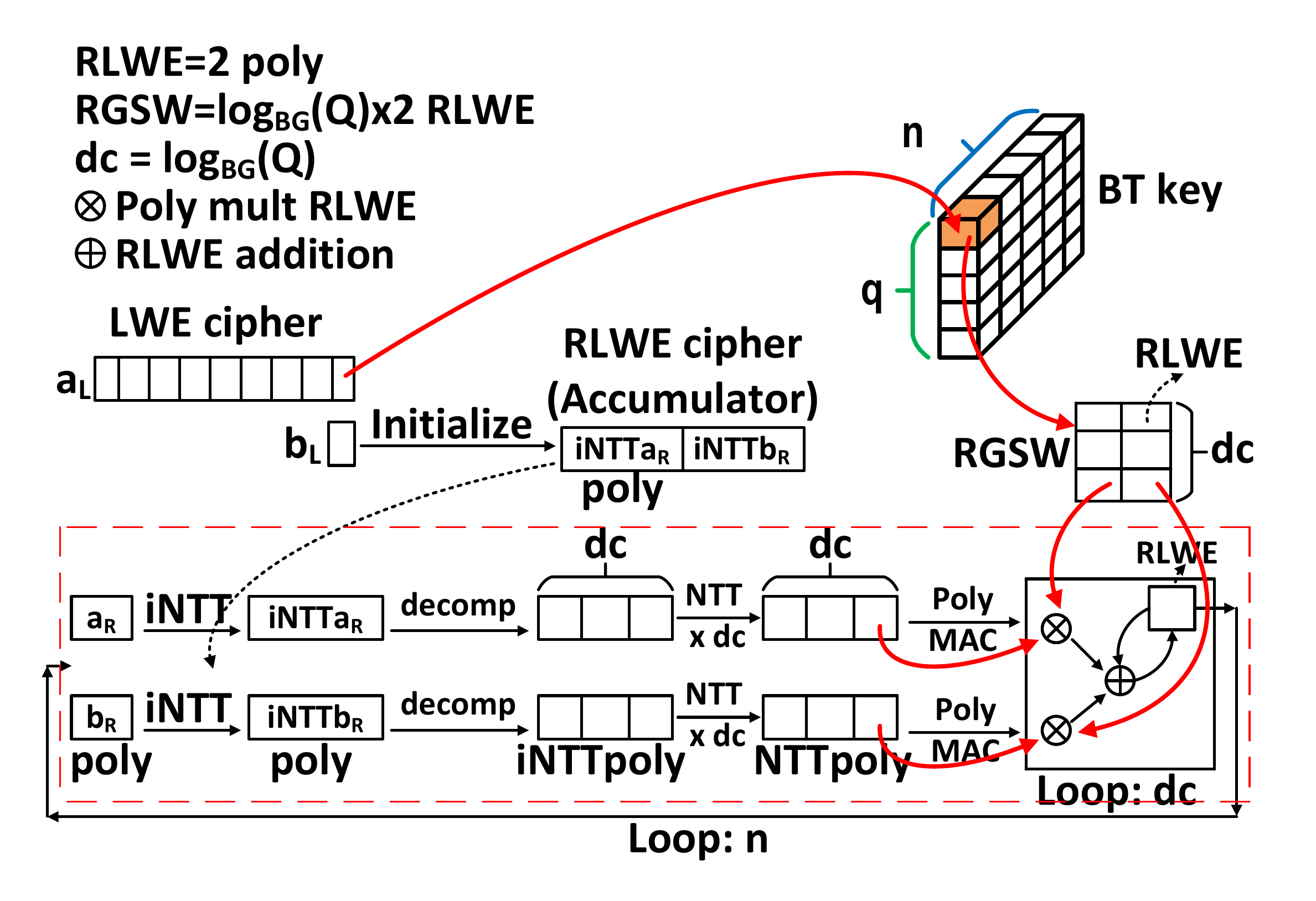}
	\centering
	\label{fig:fig4}
\end{figure}
\subsection{Bootstrap in Third-Generation FHE}
\label{sec:backgroundsub3}
As shown in Section \ref{sec:backgroundsub2}, LWE is additive homomorphic, meaning that $LWE(m_1 )+LWE(m_2 )=LWE(m_1+m_2)$, so it is used to homomorphically evaluate Boolean logic in third-generation FHE. Take $NAND(m_1,m_2)$ for example, with $m_1,m_2$ being either $0$ or $1$. The result of the NAND can be extracted from the sum $m_1+m_2+2 \mod 4$. If the sum is $2$ or $3$, then $NAND(m_1,m_2) = 0$. Otherwise, if the sum is $0$, then $NAND(m_1,m_2) = 1$. Thus, the NAND is encoded in the MSB of the sum. However, further addition cannot be applied to the resulting LWE ciphertext due to both the mismatch of the data format (LSB vs. MSB) and the increased noise, which can potentially contaminate the message. Therefore, the bootstrap process introduced in Section \ref{sec:introsub1} is required to reset the data format and noise level.

The bootstrap process of FHEW \cite{FHE10} is implemented in this work for its integer operation that better serves the purpose of hardware acceleration. It is composed of three steps, homomorphic accumulation, RLWE to LWE key switch, and LWE modulus switch. The homomorphic accumulation takes $98\%$ of the processing time \cite{FHEWlike}, therefore, this subroutine is deployed on the hardware, while others are done in software and will not be discussed here. The reader is referred to \cite{FHEWlike} for further details.

\autoref{fig:fig4} illustrates the data flow of homomorphic accumulation, with the $\bigotimes$ operation highlighted in the dotted red box. At first, the bootstrap key (BT key, an array of RGSW ciphertexts) is generated by the local user and transferred to the server. This is a one-time process. For the server to bootstrap one LWE ciphertext, a homomorphic accumulator is initialized based on $b$ of the LWE, in INTT domain. Then, the accumulator is multiplied with a element of the BT key by the $\bigotimes$ operation. The element of the BT key is indexed by $\boldsymbol{a}[i]$ of the LWE and $i\in[0,n-1]$. The product is accumulated and looped back for next multiplication. After the loop finishes, the output is passed through RLWE to LWE key switch function and LWE modulus switch function (not shown in \autoref{fig:fig4}) to complete the whole bootstrap process.

\subsection{Augmented Subroutines}
\label{sec:backgroundsub4}
To build the proposed PSI protocol, we adopt some additional features from another third-generation scheme TFHE \cite{FHE11}.

\subsubsection{$CMUX(RGSW_{\boldsymbol{z}}(\boldsymbol{m}),RLWE_{\boldsymbol{z}}(\boldsymbol{p_0}),RLWE_{\boldsymbol{z}}(\boldsymbol{p_1}))$}
The $\bigotimes$ operation between an RLWE and a RGSW, defined in Section \ref{sec:backgroundsub2sub4}, can be used to construct a homomorphic MUX gate \cite{FHE11}. Let $\boldsymbol{m}=[sel,0,0,\cdots,0]$ be the selection signal of the MUX gate with $sel$ equal to either $0$ or $1$, and let $RLWE_{\boldsymbol{z}}(\boldsymbol{p_0})$ and $RLWE_{\boldsymbol{z}} (\boldsymbol{p_1})$ be two input RLWE ciphertexts for the MUX. The CMUX function is defined in \autoref{eq:CMUX}, which output an RLWE ciphertext corresponding to the encrypted selection signal.
\begin{equation}
	\begin{split}
	&CMUX(RGSW_{\boldsymbol{z}}(\boldsymbol{m}),RLWE_{\boldsymbol{z}}(\boldsymbol{p_0}),RLWE_{\boldsymbol{z}}(\boldsymbol{p_1})) \\
	=&RGSW_{\boldsymbol{z}} (\boldsymbol{m})\bigotimes(RLWE_{\boldsymbol{z}}(\boldsymbol{p_1})-RLWE_{\boldsymbol{z}}(\boldsymbol{p_0}))+RLWE_{\boldsymbol{z}}(\boldsymbol{p_0}) \\
	=&RLWE_{\boldsymbol{z}} (\boldsymbol{m}\cdot (\boldsymbol{p_1-p_0})+\boldsymbol{p_0}) \\
	=&RLWE_{\boldsymbol{z}} (\boldsymbol{p_{sel}}) \\
	\end{split}
	\label{eq:CMUX}
\end{equation}

\subsubsection{$BlindRotate(RGSW_{\boldsymbol{z}}(\boldsymbol{m}),RLWE_{\boldsymbol{z}}(\boldsymbol{p}),j))$}
Following the definition of the $CMUX$, $BlindRotate$ is formulated to homomorphically rotate an encrypted polynomial by multiplying the polynomial with a power of $X$. A simplified version is shown in \autoref{eq:blindrotate}. Let $\boldsymbol{m}=[sel,0,0,\cdots,0]$ be the selection signal of the CMUX gate, and let $RLWE_{\boldsymbol{z}}(\boldsymbol{p})$ be the input RLWE ciphertext. Parameter $j$ denotes the number of steps for the rotation. Thus, the output RLWE encrypts a plaintext that is either rotated or not based on the selection. A comprehensive definition can be found in \cite{FHE11}.
\begin{equation}
	\begin{split}
		&BlindRotate(RGSW_{\boldsymbol{z}}(\boldsymbol{m}),RLWE_{\boldsymbol{z}}(\boldsymbol{p}),j)\\
		=&CMUX(RGSW_{\boldsymbol{z}}(\boldsymbol{m}),RLWE_{\boldsymbol{z}}(\boldsymbol{p}),X^{-j}\cdot RLWE_{\boldsymbol{z}}(\boldsymbol{p}))\\
		=&CMUX(RGSW_{\boldsymbol{z}}(\boldsymbol{m}),RLWE_{\boldsymbol{z}}(\boldsymbol{p}),RLWE_{\boldsymbol{z}}(X^{-j}\cdot\boldsymbol{p}))\\
		=&RLWE_{\boldsymbol{z}}(\boldsymbol{m}\cdot(X^{-j}\cdot\boldsymbol{p}-\boldsymbol{p})+\boldsymbol{p})\\
	\end{split}
	\label{eq:blindrotate}
\end{equation}

\begin{figure}
	\caption{(a) LUT and CMUX Tree for an Arbitrary Binary Function $f(x)$, and (b) Vertical Packing Scheme of the LUT.}
	\includegraphics[scale=0.7]{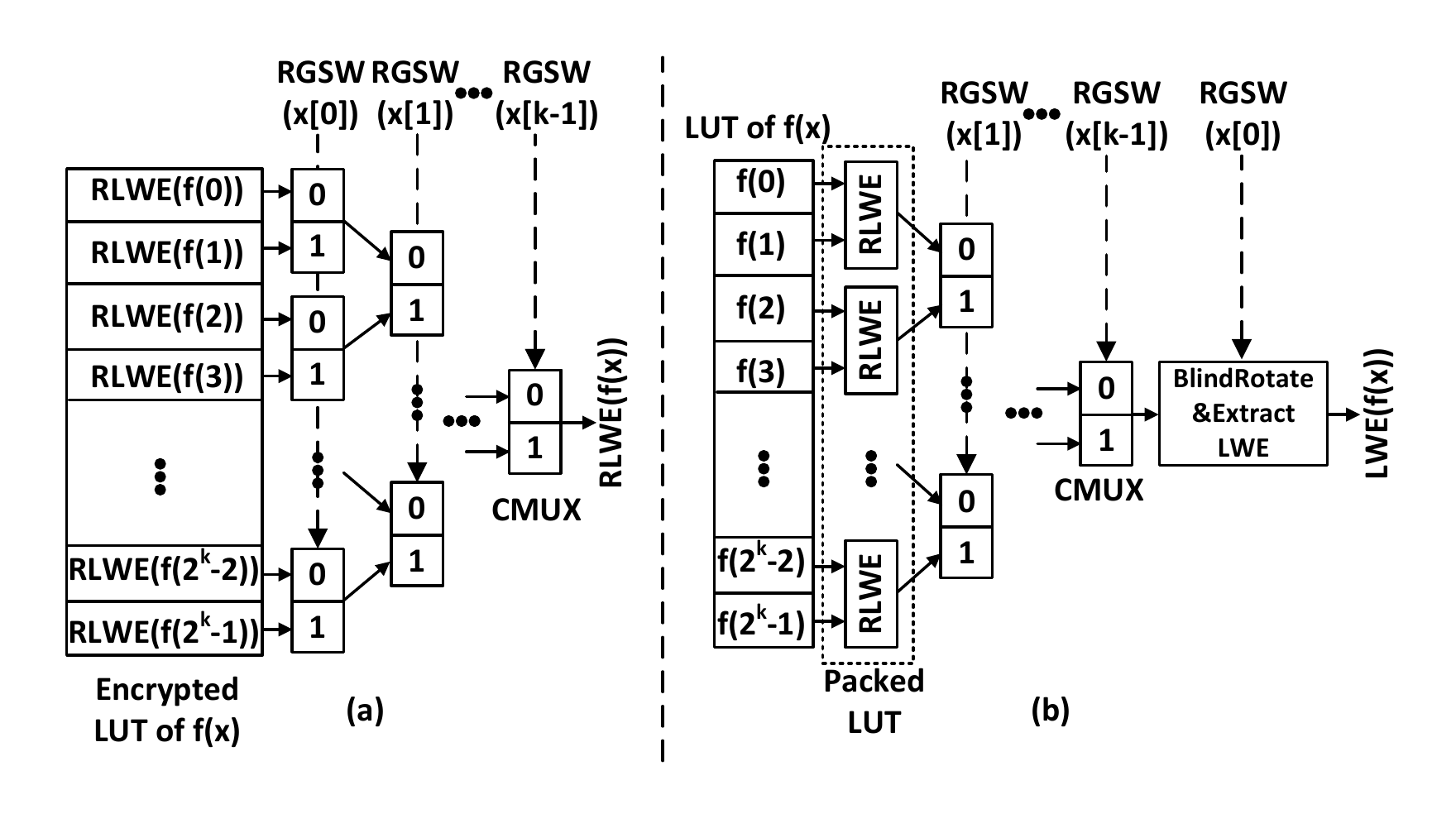}
	\centering
	\label{fig:fig78}
\end{figure}
\subsubsection{Homomorphic LUT and Plaintext Packing}
\label{sec:backgroundsub4sub3}
Intuitively, the CMUX gate can be concatenated into a CMUX tree to evaluate an arbitrary binary function homomorphically as shown in \autoref{fig:fig78} (a) \cite{FHE11}. The function is precomputed and encrypted into an LUT of RLWE ciphertexts, and after traversing the CMUX tree indexed by RGSW encryptions of the binary representation of an input $x$, an RLWE ciphertext that encrypts the corresponding $f(x)$ is output.

However, the size of the LUT is large if each RLWE ciphertext only encrypts one function value, resulting in $2^k$ RLWE ciphertexts. Also, the amount of CMUX is $2^k-1$. This exponential size can be reduced by a factor of the length of the polynomial in an RLWE ciphertext if several function values are packed into an RLWE ciphertext. For example, if each coefficient of a plaintext polynomial $\boldsymbol{m}$ is taken as a plaintext slot, then a contiguous block of function values can be packed into one polynomial, such as $\boldsymbol{m}=[f(0),f(1),\cdots,f(N-1)]$, where $N$ is the length of the polynomial. Thus, an RLWE ciphertext can encrypt at most $N$ function values, which reduces the size of the LUT and the amount of CMUX by a factor of $N$. \autoref{fig:fig78} (b) details this packing scheme using an example in which each RLWE encrypts two function values, reducing the number of CMUXs by a factor of $2$. In the example, the MSBs of the input $x$ are first used to find the desired RLWE ciphertext, and then the target slot is rotated to the position $0$ dictated by the LSB of $x$ with $BlindRotate$. Lastly, the desired slot is extracted from the RLWE into an LWE ciphertext as described in Section \ref{sec:backgroundsub2sub2}.

\subsubsection{RLWE Key Switch}
\label{sec:backgroundsub4sub4}
The last included subroutine is the RLWE key switch that converts an RLWE ciphertext encrypted under a secret key $\boldsymbol{z_1}$ into another RLWE ciphertext encrypted by a different secret key $\boldsymbol{z_2}$. Given a decomposition base $B_{KS}$, an RLWE key-switch key (KS key) is created by encrypting the secret key $\boldsymbol{z_1}$ into a vector of RLWE ciphertexts, as shown in \autoref{eq:KSvector}, with $dc=\log_{B_{KS}}Q$ denoting the length of the vector.
\begin{equation}
	[RLWE_{\boldsymbol{z_2}}(1\times \boldsymbol{z_1}),RLWE_{\boldsymbol{z_2}}(B_{KS}\times\boldsymbol{z_1} ),\cdots,RLWE_{\boldsymbol{z_2}}(B_{KS}^{dc-1}\times\boldsymbol{z_1})]
	\label{eq:KSvector}
\end{equation}

For an RLWE ciphertext $RLWE_{\boldsymbol{z_1}})(\boldsymbol{m})=[\boldsymbol{a,b}]$ encrypted with key $\boldsymbol{z_1}$, to switch to key $\boldsymbol{z_2}$, the new ciphertext is calculated by \autoref{eq:KS}, which is basically an inner product of the decomposed $a$ and the key-switch key, with $\boldsymbol{a}=\sum_{j=0}^{dc-1}\boldsymbol{a}[j]$. The multiplication of a polynomial and an RLWE ciphertext is defined in Section \ref{sec:backgroundsub2sub4}. A formal definition of the process can be found in \cite{FHE4, FHE5, FHE6}. Note that this operation resembles the $\bigotimes$ operation.
\begin{equation}
	RLWE_{\boldsymbol{z_2}}(\boldsymbol{m})=[0,b]-\sum_{j=0}^{dc-1}\boldsymbol{a}[j]\cdot RLWE_{\boldsymbol{z_2}} (B_{KS}^{j}\times\boldsymbol{z_1})
	\label{eq:KS}
\end{equation}

\begin{figure}
	\caption{(a) General Concept of Finding the Intersection of Two Sets, and (b) Homomorphic LUT Based PSI.}
	\includegraphics[scale=0.8]{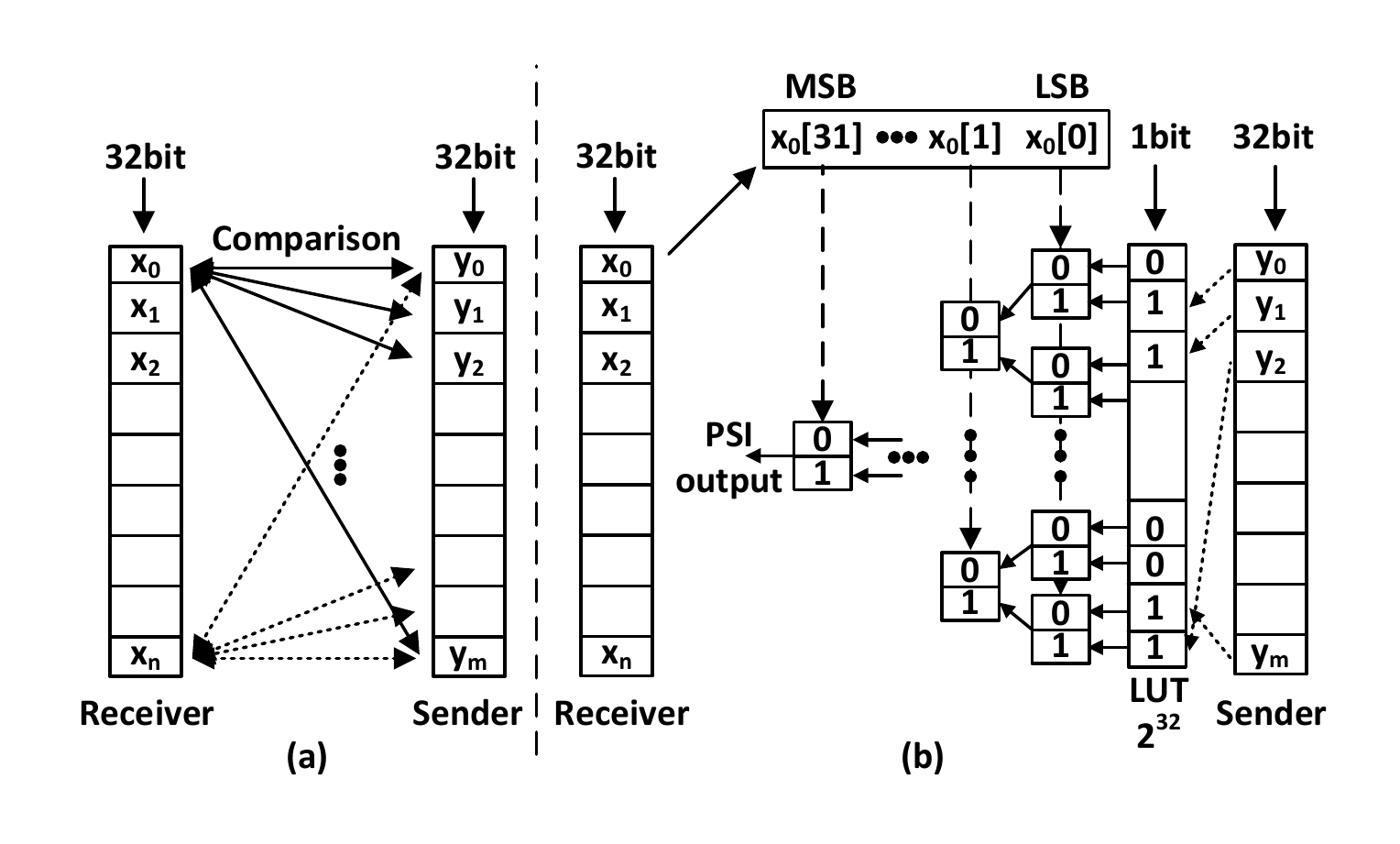}
	\centering
	\label{fig:fig910}
\end{figure}
\section{Unbalanced PSI with (Leveled) Augmented FHEW}
\label{sec:PSI}
\subsection{High Level Construction}
\label{sec:PSIsub1}
For two parties, the Receiver and the Sender, to find the intersection of their private sets $\{x_i\}$ and $\{y_j\}$ w.l.o.g. assuming each contains some 32-bit integers, as show in \autoref{fig:fig910} (a), each element of the Receiver’s set is compared with the elements of the Sender’s set. In the case of a match, the element is added to the intersection. 

However, in an unencrypted scenario, one of the parties needs to reveal all its content to the other party, which is undesirable. So, in \cite{PSI6}, the comparison is fulfilled by a homomorphic product of the difference between elements in the two sets.  For each RLWE encrypted $x_i$, the Sender evaluates homomorphically the product of the difference $y_j-x_i\forall j$, as shown in \autoref{eq:PSIprod}. After the Receiver decrypts the result, the product evaluates to $0$ if $x_i$ finds a match in the Sender’s set $\{y_j\}$. 
\begin{equation}
	\prod_{j}\big([\boldsymbol{0}, y_j]-RLWE_{\boldsymbol{z}}(x_i)\big)=RLWE_{\boldsymbol{z}}(\prod_{j}(y_j-x_i))
	\label{eq:PSIprod}
\end{equation}

In this work, the comparison is facilitated with the homomorphic LUT described in Section \ref{sec:backgroundsub4}. As shown in \autoref{fig:fig910} (b), on the Sender’s side, an LUT is precomputed based on the content of the Sender’s set $\{y_j\}$, with $LUT[y_j]=1$; otherwise, the entry is set to $0$. On the Receiver’s side, each element $x_i$ is decomposed into its binary representation and encrypted with a vector of RGSW ciphertext, $[RGSW_{\boldsymbol{z}}(x_i[0]),RGSW_{\boldsymbol{z}}(x_i[1]),\cdots,\\RGSW_{\boldsymbol{z}} (x_i[31])]$ and sent to the Sender. Then, the RGSW encrypted $x_i$ is passed into the CMUX tree to index the LUT on the Sender’s side, and the result is sent back to the Receiver. After decryption, $1$ indicates that $x_i$ is in the intersection, otherwise, it is not. Since the proposed protocol follows the high-level construction of \cite{PSI6}, the attack model, security implications, and proof in \cite{PSI6} also apply to this work with the exception that the noise flooding process is unnecessary because the LUT only returns whether $x_i$ is in set $\{y_i\}$ and therefore no excess information about set $\{y_j\}$ is leaked.

Let $b$ denote the bit width of the elements inside both sets. The communication complexity is linearly dependent on $b$ and the size of the Receiver’s set, resulting in $O(b\times|\{x_i\}|)$. The computation cost is $O(2^b)$, which is independent of the size of the Sender’s set. So, this naïve construction is very inefficient in both computation and communication traffic. For example, if $b=32$, 32 RGSW ciphertexts have to be transferred for each element in the Receiver’s set, resulting in a low ciphertext utilization. Additionally, $2^{32}-1$ CMUXs are evaluated for each element in the Receiver’s set. Several optimizations can be adopted to mitigate these problems and render practical application of the protocol. 

\begin{figure}
	\caption{Data Flow of the RLWE Substitution Subroutine (RLWE key Switch Included).}
	\includegraphics[scale=0.5]{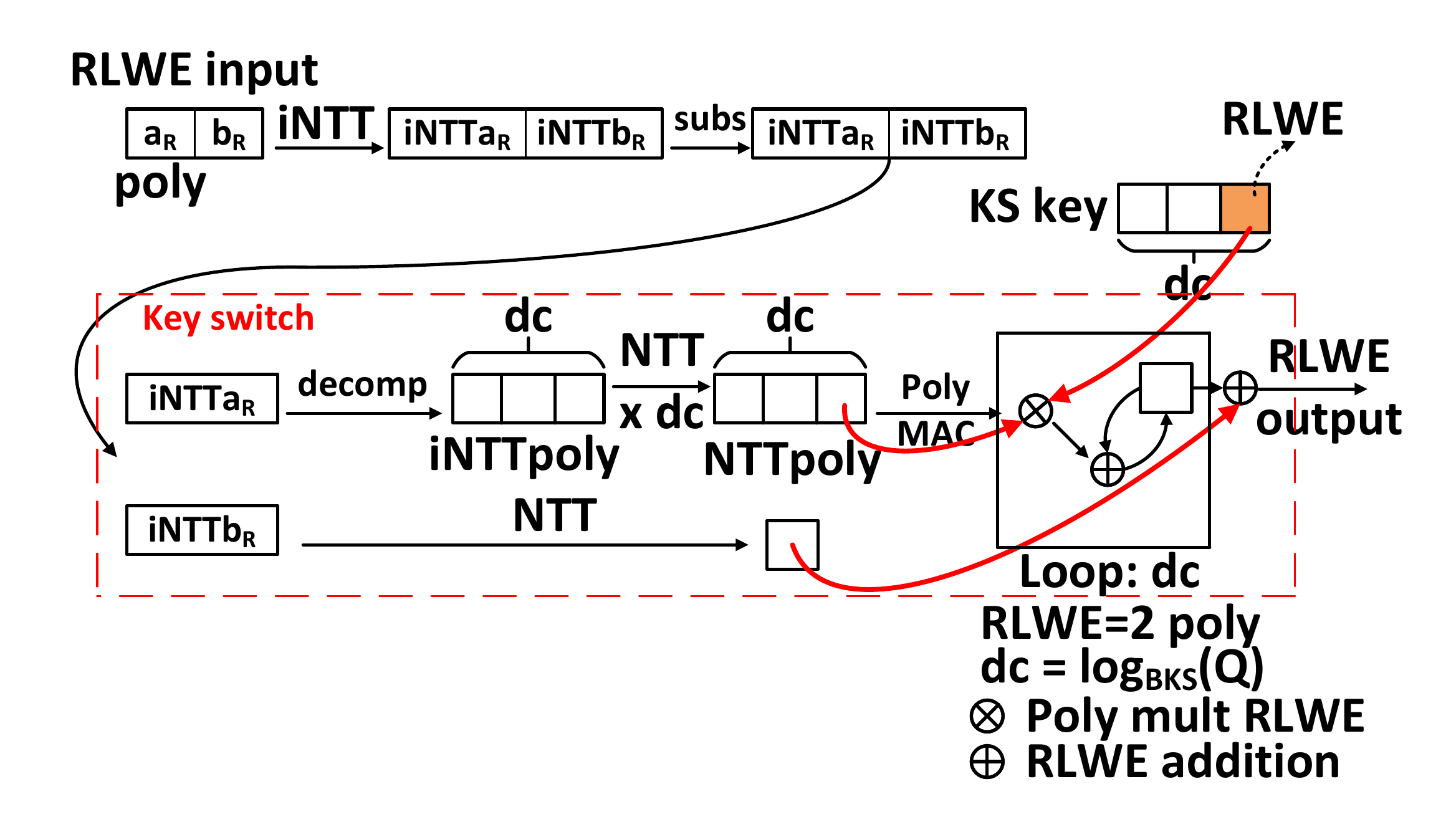}
	\centering
	\label{fig:fig11}
\end{figure}

\subsection{RLWE Substitution and RLWE Expansion}
\label{sec:PSIsub2}
Before tackling the problems, two additional subroutines need to be discussed. The first is RLWE substitution, which transforms an RLWE ciphertext $RLWE_{\boldsymbol{z}}(\sum\boldsymbol{m}[i]X^i)$ into $RLWE_{\boldsymbol{z}} (\sum\boldsymbol{m}[i](X^i)^k)$ for an odd integer $k$. An RLWE key-switch key from $\boldsymbol{z}(X^k)$ to $\boldsymbol{z}$ is precomputed based on the substituted secret key $\boldsymbol{z}(X^k)=\boldsymbol{z}[i] (X^i)^k$. In the process, an RLWE ciphertext is first substituted to get $RLWE_{\boldsymbol{z}(X^k)}\big(\boldsymbol{m}(X^k)\big)=[\boldsymbol{a}(X^k ),\boldsymbol{b}(X^k)]$, and then key-switched to $RLWE_{\boldsymbol{z}}(\boldsymbol{m}(X^k))$ encrypted with the original secret key. A formal definition can be found in \cite{ORAM2}.

The RLWE substitution is used extensively in the RLWE expansion operation \cite{ORAM2}, which expands an RLWE ciphertext from $RLWE_{\boldsymbol{z}}(\sum\boldsymbol{m}[i] X^i)$ into a vector $[RLWE_{\boldsymbol{z}}(\boldsymbol{m}[0]),\allowbreak RLWE_{\boldsymbol{z}} (\boldsymbol{m}[1]),\allowbreak\cdots,RLWE_{\boldsymbol{z}}(\boldsymbol{m}[N-1])]$. An example of how RLWE substitution fulfills the expansion is detailed in Appendix \ref{sec:appRLWEexpan}.

The data flow of RLWE substitution is shown in \autoref{fig:fig11}, with the key switch highlighted in the dotted red box. An RLWE ciphertext in the NTT domain is first transformed into INTT form and substituted. Then it is decomposed with base $B_{KS}$ and key-switched to the original secret key to get a substituted RLWE ciphertext. After that, the output ciphertext is postprocessed for RLWE expansion. Based on our experiment, $97\%$ of the processing time of RLWE expansion is dedicated to substitution and key switch functions, so these two functions are offloaded to a FPGA. Note that the key switch data flow is very similar to the bootstrap data flow. Therefore, the proposed architecture merges both data flows, which will be detailed in Section \ref{sec:ACCArch}.

\subsection{Optimizations for The PSI Protocol}
\label{sec:PSIsub3}
For the proposed PSI, both the computation and communication costs depend directly on the bit width of the elements in the set. Hence, the first optimization is to reduce the bit width of the element with permutation-based hashing \cite{permhash}. In permutation-based hashing, to insert a 32-bit element $x_i$ from the Receiver’s set into $2^k$ bins, it is divided into $x_{iH}||x_{iL}$, with $x_{iL}$ consisting of $k$ bits. The position of the element is calculated by \autoref{eq:permhash}, where $H(x)$ is a hash function. Therefore, the position of an element also stores some information about the element. And instead of inserting $x_i$ into the hash table, only $x_{iH}$ is inserted, which reduces the bit width to $32-k$.The correctness of the comparison in the homomorphic LUT holds with permutation-based hashing, which is detailed in Appendix \ref{sec:appPermhash}. With permutation-based hashing, the amount of transferred RGSWs is reduced by $k$ and the amount of CMUXes is reduced by a factor of $2^k$.
\begin{equation}
	pos(x_i)=H(x_{iH})\ XOR\ x_{iL}
	\label{eq:permhash}
\end{equation}

The second optimization achieves further computational reduction by exploiting the vertical packing described in Section \ref{sec:backgroundsub4}. With vertical packing, at most $N$ LUT elements can be packed into one RLWE ciphertext, which shrinks the amount of CMUXs by roughly a factor of $N$. For example, after the permutation hashing with $k=14$, the bit width of the elements in each bin is $18$, reducing the size of the CMUX tree to $2^{18}-1$. At $N=2048$, the vertical packing reduces it to $2^7-1+11$. Compared to the original size, $2^{32}-1$, a reduction by $2^{25}$ times is achieved in total.

The last optimization aims at decreasing the communication payload. Instead of transferring an RGSW ciphertext, containing $2\log_{B_G}(Q)$ RLWE ciphertexts, for each bit of an element in the Receiver’s set, it is observed that the first column of an RGSW ciphertext, as shown in \autoref{eq:RGSW} and Section \autoref{eq:RLWEPrime}, can be calculated from the second column, which is detailed in \autoref{eq:RGSWseccol}. Thus, only the second column needs to be transferred together with a shared RGSW encryption of the secret key $-\boldsymbol{z}$ \cite{ORAM2}. The transaction size is therefore reduced by a factor of $2$.
\begin{equation}
	RLWE_{\boldsymbol{z}}(B_G^j \times(-\boldsymbol{z\cdot m}))=RLWE_{\boldsymbol{z}} (B_G^j\times m)\bigotimes RGSW_{\boldsymbol{z}} (-\boldsymbol{z})
	\label{eq:RGSWseccol}
\end{equation}

However, the ciphertext utilization is still very low because for each element in the Receiver’s set, $\log_{B_G}Q$ RLWEs are transferred. So, $N$ elements, for example, $[x_0,x_1,\cdots,x_{N-1}]$, from the Receiver’s set are packed into a 2-D array of RLWE ciphertexts for better utilization. Each element of the array is formed as $RLWE_{\boldsymbol{z}}(\sum_{i}x_i [k] \times B_G^j \times X^i)$ and is indexed by $j\in[0,log_{B_G}(Q)-1]$, $k\in[0,17]$ (assuming after applying permutation-based hashing, the bit width is $18$). Upon receiving the array, the Sender unpacks it, with the RLWE expansion described in Section \ref{sec:PSIsub2}, into arrays of RLWEs for each element $x_i$, of the form $RLWE_{\boldsymbol{z}} (x_i [k]\times B_G^j )$. Finally, the RLWEs are converted into RGSWs with \autoref{eq:RGSWseccol} and passed into the LUT to complete the PSI. Note that we set the $B_G$ of the RGSW to be equal to the $B_{KS}$ of the key-switch key used in the RLWE expansion. 

Together, compared to transferring complete RGSWs, the communication overhead is reduced from $18$ RGSWs ($216$ RLWEs) per element to $18\times\log_{B_G}(Q)$ RLWEs per $N$ elements, amounting to a $4096\times$ reduction if  $\log_{B_G}(Q)=6$, $N=2048$, at the cost of increased computation on the Sender’s side to unpack and reconstruct the RGSWs.

\begin{figure}
	\caption{Data Flow of the Homomorphic LUT Based PSI.}
	\includegraphics[scale=0.34]{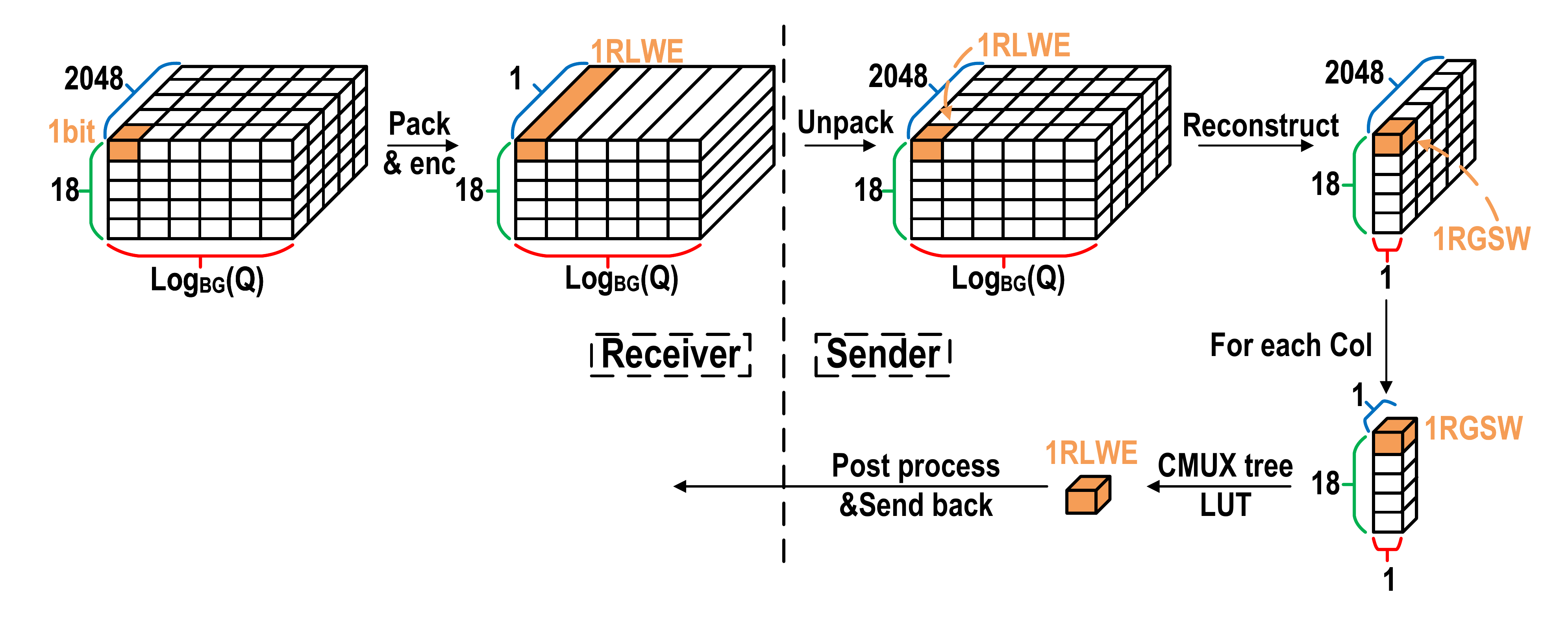}
	\centering
	\label{fig:fig12}
\end{figure}

\autoref{fig:fig12} shows the data flow of the proposed homomorphic LUT-based PSI. It assumes that after the permutation-based hashing, the data bit width is $18$ bits and the polynomial length is $N=2048$. The Receiver packs all the necessary bits into an array of RLWE ciphertexts and sends it to the Sender. The Sender then unpacks and the array of RLWEs into an array of RGSWs in which each column encrypts the binary representation of an element in the Receiver’s set. Then, each column of RGSW is passed through the LUT, and an RLWE that encrypts the lookup result at index 0 is generated. Finally, the LWEs that encrypt lookup results are extracted from the RLWEs, as described in Section \ref{sec:backgroundsub2}, and sent back to the Receiver. The hashing process is not shown in the figure. Other optimizations utilized in \cite{PSI6} can also be applied to our protocol, such as pre-hashing both parties’ sets into smaller sets to reduce the set sizes, using modulus switching to reduce reply ciphertext size, etc.

In summary, after applying the above optimizations, the communication overhead of the scheme is $O(\frac{b-k}{2N}\times2^k)$, assuming, on the Receiver’s side, $2^k$ bins after hashing and at most one element in each bin, with dummy elements filling up the empty bins. The computation cost is $O(\frac{2^b}{N}+2^k (b-k))$. 

\begin{figure}
	\caption{(a) Overall Architecture of the Proposed Accelerator, and (b) Overview of the Main Compute Chain.}
	\includegraphics[scale=0.45]{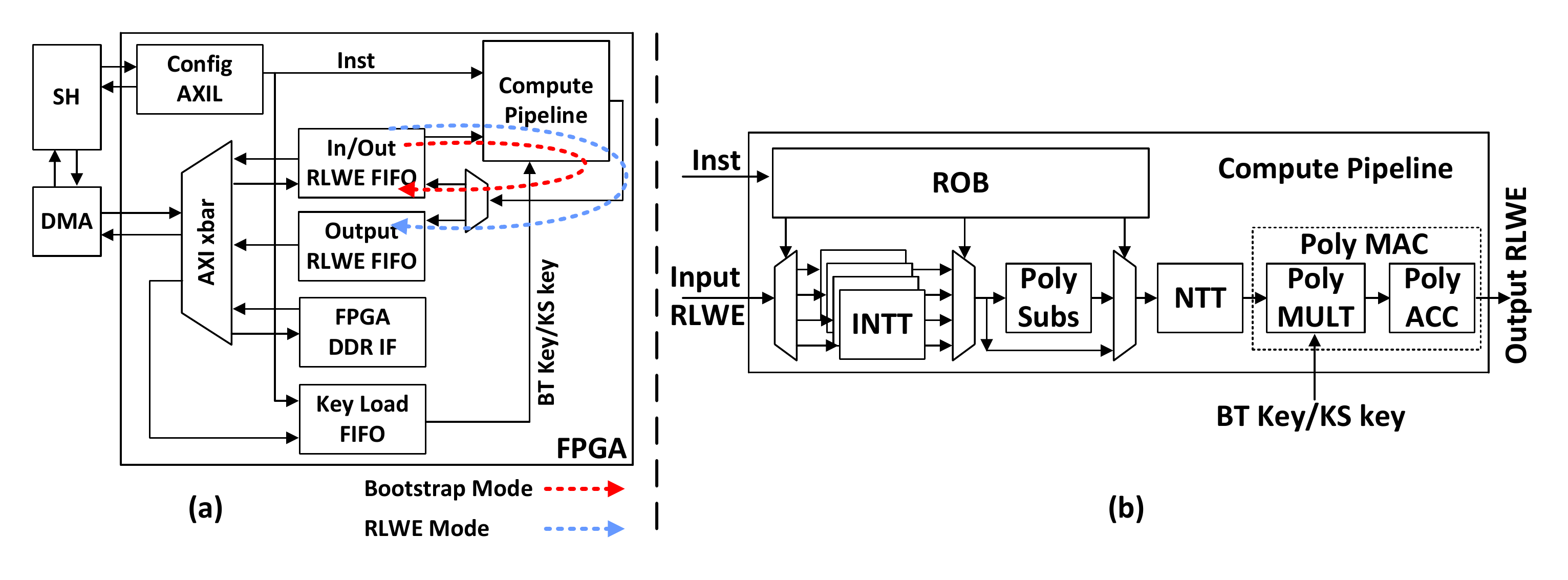}
	\centering
	\label{fig:fig1314}
\end{figure}

\section{Architecture of The Proposed Accelerator}
\label{sec:ACCArch}

\subsection{Overall Architecture}
\label{sec:ACCArchsub1}
\autoref{fig:fig1314} (a) shows the overall architecture of the proposed accelerator with a zoom-in view of the compute pipeline in \autoref{fig:fig1314} (b). Implemented with AWS F1 instance, the accelerator is controlled and monitored by the host software running on an x86 processor through various AXI interfaces. The configure parameters and instructions are programed with the AXI-Lite interface, and the FIFO states are also read from it. The DMA module communicates with the FPGA through the AXI bus to program the FPGA DDR and read/write the RLWE FIFOs. The RLWEs streamed in and out of the FPGA are in the NTT domain. Further, the modulo multiplication in the accelerator is facilitated by the standard Barrett Reduction \cite{Barrettreduction}.

The architecture works in a pipelined fashion, with necessary inter-stage double buffering. Upon an input instruction, the key load module reads the corresponding key from the preprogramed FPGA DRAM into its own key load FIFO. In parallel, the INTT/NTT modules inside the compute pipeline manipulate the input RLWEs, hiding the DDR access delay of the key load module since the keys are only needed at the poly MAC stage, which facilitates the polynomial and RLWE vector inner product introduced in Section \ref{sec:backgroundsub2sub4} and Section \ref{sec:PSIsub2}. Once the computation finishes, the output RLWEs are written back to the RLWE FIFO dictated by the mode of the accelerator, which will be detailed later, and then streamed out to the host. 

As mentioned in Section \ref{sec:PSIsub2}, the accelerator merges the two data flows, the RLWE substitution and the bootstrap process. Note that the data flow of evaluating the homomorphic LUT introduced in Section \ref{sec:backgroundsub4sub3} is mostly the same as the bootstrap flow since they both incorporate the $\bigotimes$ operation, so they will not be differentiated in the remaining text. There are three primary differences between the two data flows. The first is the RLWE key switch versus the $RLWE\bigotimes RGSW$ operation as highlighted in \autoref{fig:fig11} and \autoref{fig:fig4}, respectively. Second, in RLWE substitution, after INTT, the subroutine that transforms the $RLWE_{\boldsymbol{z}}(\boldsymbol{m})$ into $RLWE_{\boldsymbol{z}(X^k)} (\boldsymbol{m}(X^k))$, as stated in Section \ref{sec:PSIsub2}, is needed; this subroutine is unnecessary in the bootstrap process. Lastly, an RLWE ciphertext, streamed into the in/out FIFO, only passes through the compute pipeline once for RLWE substitution and is then streamed out from the output FIFO after the computation. In contrast, in the bootstrap process, after initialization, the same RLWE (homomorphic accumulator) must be looped $n$ times through the compute pipeline before being streamed out, meaning that the output RLWE from the compute pipeline should go to the same FIFO as the input RLWE. 

The first two differences regarding the computation are automatically taken care of by the different instructions passed into the compute pipeline. For the third one, a mode configuration is added to the FIFOs to differentiate the situations, as shown by the dotted lines in \autoref{fig:fig1314} (a). In $RLWE$ mode, the in/out FIFO acts only as an input FIFO that receives the input RLWEs, whereas the output FIFO holds the processed RLWEs. While in the $bootstrap$ mode, the output FIFO is turned off and the in/out FIFO holds the intermediate RLWEs. The compute pipeline continuously reads and writes the in/out FIFO until the loop finishes. Then the RLWEs in the FIFO are streamed out to the host.

\begin{figure}
	\caption{(a) Architecture and Dataflow of the INTT Module, (b) the Time-Interleaving of the Polynomial Buffers, and (c) Two Data Access Pattern of The INTT Module.}
	\includegraphics[scale=0.8]{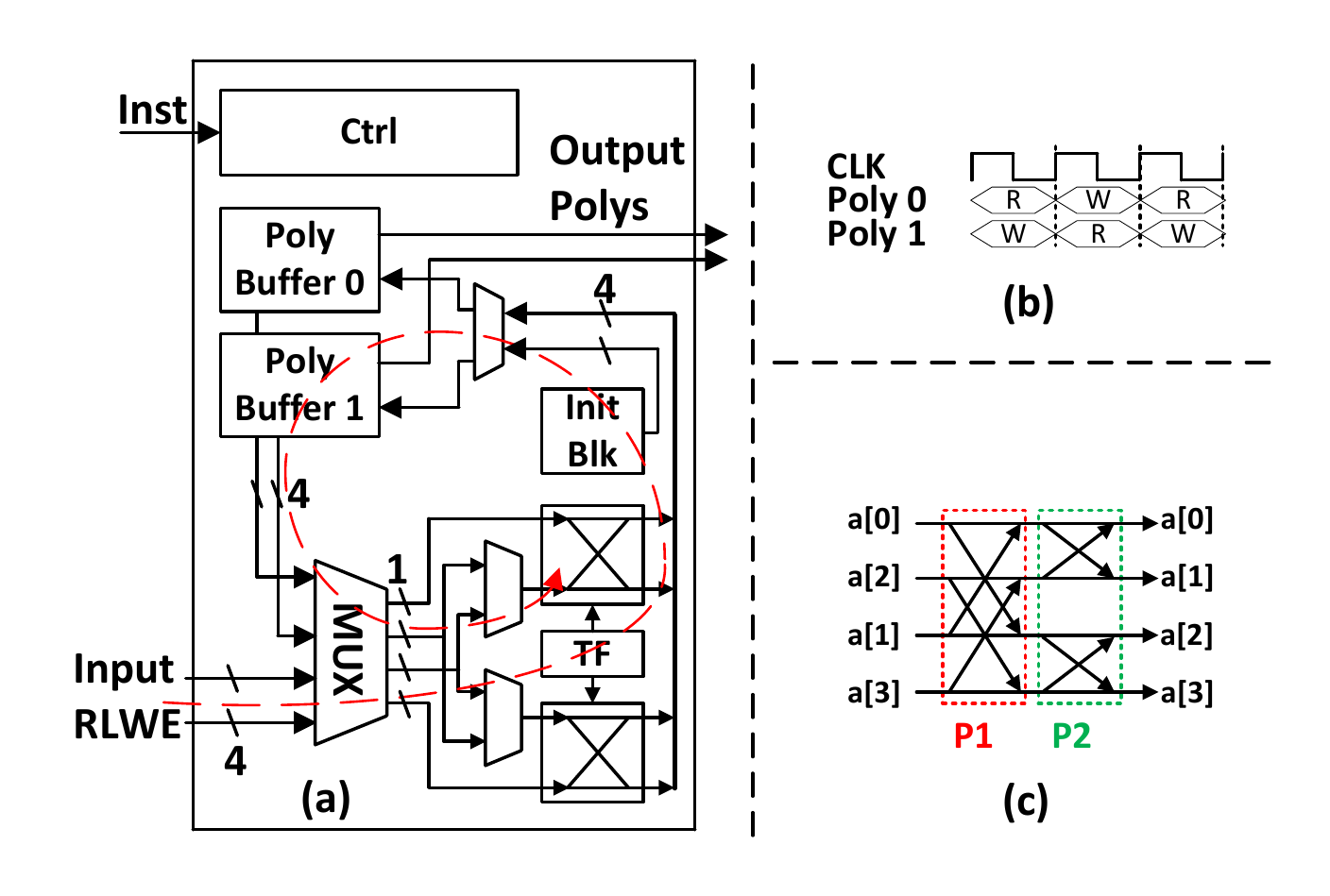}
	\centering
	\label{fig:fig1516}
\end{figure}

\subsection{INTT Module}
\label{sec:ACCArchsub2}
\autoref{fig:fig1516} (a) details the structure of the INTT module, which follows the algorithm in Appendix \ref{sec:appINTT}, except for the first outer loop where the input RLWE is read from the global in/out RLWE FIFO, and the intermediate result is written to its own two polynomial buffers since each RLWE contains two polynomials. Starting from the second outer loop, the input is read from the polynomial buffers and written back after being processed by the butterflies. Each INTT module stores its own copy of the twiddle factors (TFs) in its local memory.

Two parallel butterfly units are included in each INTT module to achieve better performance while maintaining a reasonable FPGA resource usage. Thus, to feed enough data, each address of the polynomial buffer contains two consecutive coefficients of a polynomial. The BRAMs of the FPGA that are used to build the polynomial buffers are inherently composed of two read/write ports, fitting the butterfly data access pattern and allowing it to read/write two different addresses at the same time. However, the read and write can only be done in separate clock cycles, resulting in $50\%$ butterfly utilization and halving the throughput. Therefore, we time interleave the two polynomial buffers, as shown in \autoref{fig:fig1516} (b), to achieve full utilization of the butterflies.

Due to the variation of the data access pattern of the butterfly units in each outer loop of the INTT algorithm, there is a mismatch between the data access pattern and data storage pattern, resulting in two different data flows from the buffers to the butterflies. As shown in \autoref{fig:fig1516} (c), in pattern 1, the data passes into a butterfly are from different addresses, while in pattern 2, they are from the same address. Therefore, necessary data MUXs are appended to the butterfly units to reorder the input/output data as needed. All the necessary loop counters and step counters are implemented inside the control block, together with the control of the MUXs. 

Besides the INTT functionality, the INTT module also incorporates an init block for the homomorphic accumulator initialization function mentioned in Section \ref{sec:backgroundsub3}.

\subsection{Pipelined NTT Module}
\label{sec:ACCArchsub3}
\begin{figure}
	\caption{Architecture and Dataflow of the Pipelined NTT Module.}
	\includegraphics[scale=0.7]{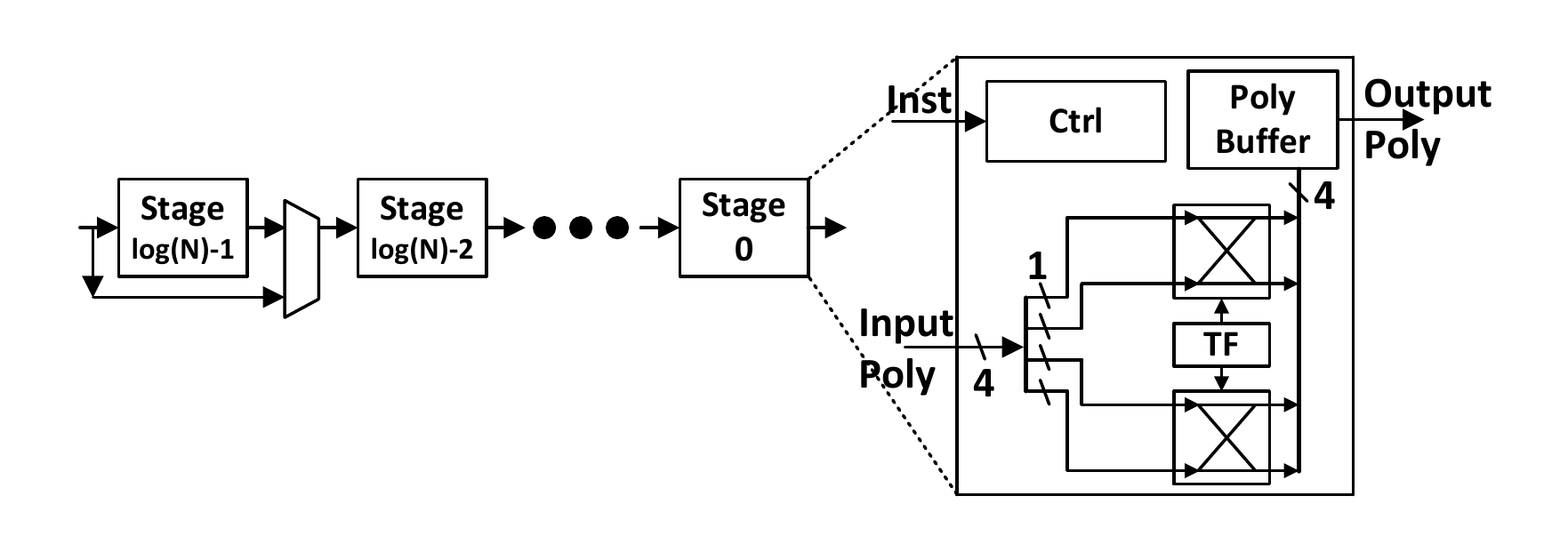}
	\centering
	\label{fig:fig17}
\end{figure}

The NTT algorithm (Appendix \ref{sec:appNTT}) is very similar to the INTT algorithm, except for the last scaling loop \cite{NTTTRICK}. But a different construction from the INTT module is adopted for the NTT module. The structure of the module is shown in \autoref{fig:fig17}, and a discussion of this construction is included in a later section. 

To achieve higher throughput for the NTT module, the outer loop of the NTT algorithm is unrolled into $\log(N)$ pipeline stages, with each stage only processing one fixed data access pattern, greatly reducing the control complexity of each stage. Compared to the structure of the INTT module, this implementation offers the same processing latency for an input polynomial but $\log(N)$ times higher throughput.

Each stage reads the input from the polynomial buffer of the previous stage and processes it with a predetermined data access pattern that is specific to that stage at design time. So, there is no on-the-fly control/MUXs for the data flow, which not only reduces resource usage but also allows a better timing requirement. Note that there is no read/write from/to the same buffer memory; therefore, it is not necessary to employ the time-interleave trick as in the INTT module. 

Although the internal structures of the stages are mostly the same, except for the loop counter and step counter inside the control block, extra care should be taken in actual implementation. First, to adapt to different polynomial lengths, MUXs are needed to skip the leading stages for short polynomials (\autoref{fig:fig17}). Second, the leading stages also incorporate the decomposition functionality as stated in Section \ref{sec:backgroundsub2sub4} and Section \ref{sec:backgroundsub4sub4}, which is just a bitwise AND with a binary decomposition basis and is not detailed in the figure.

\begin{figure}
	\caption{Comparison of Symmetric (a) and Asymmetric (b) Compute Pipelines.}
	\includegraphics[scale=0.5]{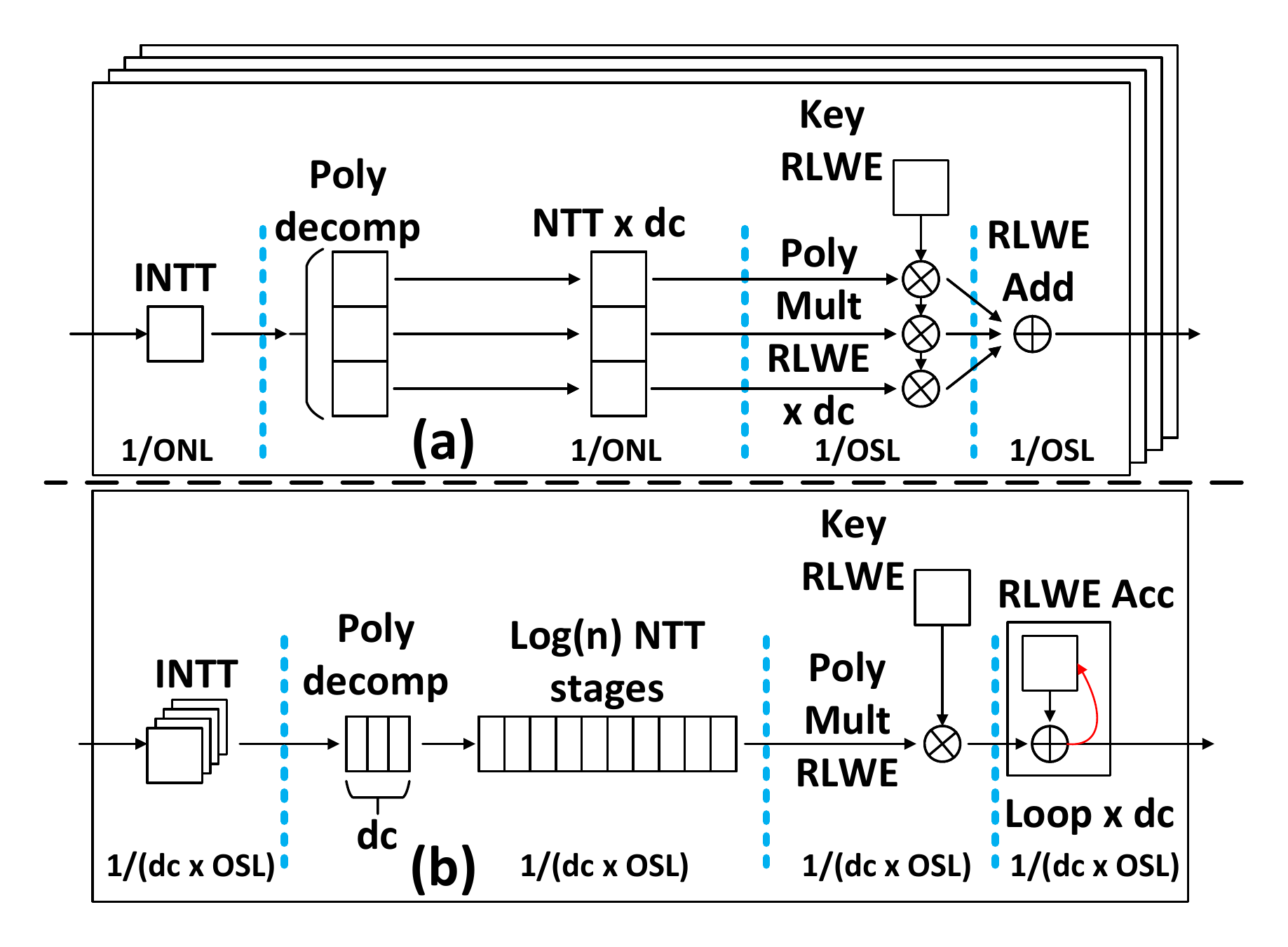}
	\centering
	\label{fig:fig18}
\end{figure}

\subsection{Compute Pipeline Analysis: Asymmetric INTT and NTT}
\label{sec:ACCArchsub4}
In the following section, we refer to the overall latency of the INTT/NTT algorithms as one NTT latency (ONL) and the latency of one outer loop of the algorithm as one stage latency (OSL). Therefore, $ONL=\log(N)\times OSL$. And our compute pipeline architecture, utilizing the pipelined NTT module (\autoref{fig:fig17}) with non-pipelined INTT modules, is referred to as asymmetric structure due to the throughput difference of the two styles. The conventional implementation of using similar structures and latencies for both NTT and INTT modules (\autoref{fig:fig1516}) is defined as a symmetric structure.

The design of our compute pipeline concentrates on balancing high throughput with optimized resource usage and parameter flexibility. So, the main compute pipeline is built around an asymmetric structure, as shown in \autoref{fig:fig1314} (b). A comparison of the symmetric and asymmetric structures is given in \autoref{fig:fig18}, with the poly subs block omitted as it is not a throughput bottleneck. The dataflows of the $RLWE\bigotimes RGSW$ (\autoref{fig:fig4}) and RLWE substitution (\autoref{fig:fig11}) can be mapped to both architectures with the same throughput. However, the asymmetric structure consumes less resources than its symmetric counterpart. 

In the symmetric pipeline (\autoref{fig:fig18} (a)), to have balanced throughput, one INTT module is accompanied with $dc$-many NTT modules since each input polynomial is decomposed into $dc$ polynomials after the INTT operation. The throughput of both modules is one polynomial per ONL due to the non-pipelined construction. The NTTs are also followed by $dc$-many polynomial/RLWE multiplication blocks to facilitate the inner product of the two dataflows. Although the trailing stages can operate with higher throughput, the overall throughput of the whole pipeline is capped by the first two stages, resulting in a throughput of one polynomial per ONL. Higher throughput can be achieved by operating multiple pipeline instances in parallel.

Most of the prior arts implemented an architecture that is similar to the symmetric structure with the INTT and NTT modules separated without considering the data flow connecting the modules. We take it one level up and make use of the asymmetric structure to cope with the different throughput requirements of the INTT and NTT modules, as shown in \autoref{fig:fig18} (b). Since the throughput of the pipelined NTT is OSL, the overall throughput of the whole pipeline is one input polynomial per $dc\times OSL$ because of the polynomial decomposition, with one caveat that to balance the throughput between the INTT and NTT, $\log(N)/dc$ INTT modules should operate in parallel. Note that in the asymmetric scenario, the trailing stages are also changed to the pipelined form ($1$ vs. $dc$ poly mult RLWE and an accumulation vs. a wide addition). In practice, $\log(N)$ is always greater than $dc$; therefore, the asymmetric pipeline enables higher throughput than a single instance of the symmetric pipeline.

Though the symmetric structure can achieve the same throughput as the asymmetric one, with $\log(N)/dc$-many instances operating in parallel, as seen in \autoref{fig:fig18} (a), the asymmetric pipeline uses less FPGA resources. The reduced resources stem from three sources. First, it is clear that in both cases, the number of INTT modules is the same, amounting to $\log(N)/dc$. The number of NTT modules seems to be the same as well since there are $\log(N)/dc\times dc =\log(N)$ NTT modules for the symmetric pipeline and the asymmetric one also incorporates $\log(N)$ NTT stages. However, in the symmetric case, the NTT module has a similar structure as the INTT module shown in \autoref{fig:fig1516} (a), which is much more complex than the NTT stage used in the pipelined NTT. Synthesis shows that with pipelined NTT, $23\%$ less LUT usage is achieved.

Furthermore, the pipelined NTT module has not only smaller control logic but also lower memory requirements. Part of the savings comes from less TF memory in pipelined NTT. Each of the NTT modules used in the symmetric pipeline stores a complete copy of the polynomial of the TF in its own local memory, similar to what is shown in \autoref{fig:fig1516} (a), so that they can operate independently. Therefore, in total, $\log(N)$ copies of the TF are stored. In contrast, in the asymmetric version, there is only one complete copy of the TF. Because each stage of the pipelined NTT is only responsible for one outer loop of the NTT algorithm, it only needs to store the portion of the TF that is used in that outer loop. For example, in the first stage of the pipelined NTT, instead of a complete polynomial of TF with $N$ coefficients, only one TF needs to be stored. Thus, overall, a $\log(N)$ times reduction of the TF memory usage is achieved with pipelined NTT, equivalent to over $10\times$ reduction in practice. It is possible to reduce the memory usage in the symmetric version by sharing one TF memory within one pipeline and force all the NTT modules to act at the same pace, but that implies stricter timing requirements since the capacitive load of the memory output is $dc$ times higher, exacerbating performance. Also forcing all NTTs to synchronize degrades the flexibility of the architecture.

The memory size of the pipelined NTT module is also reduced due to fewer polynomial buffers. In the non-pipelined NTT module, similar to the INTT module in Section \ref{sec:ACCArchsub2}, two polynomial buffers are instantiated for time-interleaved buffer access to maintain $100\%$ butterfly utilization. In contrast, each stage of the pipelined NTT module reads and writes different buffers; therefore, time-interleaving is unnecessary. So, the pipelined NTT poses a $50\%$ saving on the polynomial buffer compared to the non-pipelined version. 

Lastly, the trailing stages of the asymmetric pipeline are also less complex than that of its symmetric counterpart. As shown in \autoref{fig:fig18}, since the pipelined NTT outputs one polynomial at a time, only one poly mult RLWE module is needed in the asymmetric structure, compared to $\log(N)$ parallel mult modules in the symmetric one. In practice, it reduces the amount of mult modules by $11$ times, with $N=2048$. Although the amount of poly mult RLWE modules can be reduced in the symmetric pipeline by reusing one mult module across different NTTs in a time-interleaved manner due to the higher throughput compared to the INTT/NTT, a very wide MUX, $\log(N)$ to one, must be inserted between the stages, which would greatly impact timing and performance and introduce more control complexity. In the asymmetric structure, there is a similar MUX between the INTT and NTT modules; however, it is only $\log(N)/dc$ to one, which is much smaller. In addition, the $dc$-wide RLWE addition in the symmetric pipeline is also replaced with an RLWE accumulation with ordinary word-size modulo addition.

Besides the resource savings, the asymmetric construction also automatically adjusts to different parameter settings. In the symmetric pipeline, the number of NTTs should be set as the largest possible number of dc of the application at design time. If at design time, the parallelism is 3 for NTT, when $dc = 2$ at run time, the utilization of the NTTs is only $66.7\%$. Extra effort can be applied to remap the connection between INTT and NTT to reach $100\%$ utilization, but that comes with more control overhead, negatively impacting performance. However, with the pipelined NTT module, as long as the INTT continuously feeds input to it, $100\%$ utilization is always maintained with no extra control overhead involved since the design space of the pipelined NTT itself is independent of the parameter $dc$. In fact, even when the $dc$ of run time is higher than the designated $dc$ of design time, the pipelined NTT requires no extra control to handle it. However, it should be noted that in the above case, the INTT of the asymmetric pipeline can be underutilized. But since the number of INTT blocks is less than the number of NTT stages in general, it is not optimized in this work. 

\section{Measurement}
\label{sec:measurement}
\subsection{Experiment Setup}
\label{sec:measurementsub1}
The proposed architecture is implemented at 125MHz system frequency on an AWS F1 instance. The implementation supports up to $54$-bit input data word size. But to reduce the complexity of the modulo multiplication block in the butterfly, only a subset of the bit widths is implemented,as detailed in the following sections. Two polynomial lengths, $1024$ and $2048$, which are typical for third-generation FHE and fit our experiment for the PSI protocol, are supported natively. The polynomial buffers in the FIFOs, implemented with BRAM, are configured to the size of the longer length, $2048$. In addition, since $2$ butterfly units operate at the same time in the INTT and NTT, each buffer line contains two consecutive polynomial coefficients. Thus, the size of each polynomial buffer is predefined as $1024\times108$ bit. However, in this prototype, no optimization on the BRAM utilization is devised, so when the input polynomial length is $1024$, only the first half of the buffer is used. Following the analysis of Section \ref{sec:ACCArchsub4}, the number of INTT is set to largest possible $\log(N)/dc$ to keep a balanced throughput, which is $4$ in our implementation.

\begin{table}

	\caption{Parameter Sets of The Third Generation FHE}	
	\begin{center}
		\begin{tabular}{ |c|c|c|c|c|c|c|c| } 
			\hline
			Parameter Set & $n$ & $q$ & $N$ & $\log_2(Q)$ & $B_{ks}$ & $B_G$ & $B_r$\\
			\hline
			MEDIUM	& 256 & 512 & 1024 & 27 & 25 & $2^9$ & 23 \\
			\hline
			STD128\_AP & 512 & 512 & 1024 & 27 & 25 & $2^9$ & 23 \\
			\hline
			STD192 & 512 & 512 & 2048 & 37 & 25 & $2^{13}$ & 23 \\
			\hline
			STD256 & 1024 & 1024 & 2048 & 29 & 25 & $2^{10}$ & 32 \\
			\hline
			STD192Q & 1024 & 1024 & 2048 & 35 & 25 & $2^{12}$ & 32 \\
			\hline
			STD256Q & 1024 & 1024 & 2048 & 27 & 25 & $2^7$ & 32 \\
			\hline
		\end{tabular}
	\end{center}
	\label{tab:tb1}
\end{table}
\begin{table}
	\caption{Measured Processing Time of Homomorphic Accumulation Compared to Software}	
	\begin{center}
		\begin{tabular}{ |>{\centering\arraybackslash}p{6em}
						|>{\centering\arraybackslash}p{6em}
						|>{\centering\arraybackslash}p{6em}
						|>{\centering\arraybackslash}p{6em}
						|>{\centering\arraybackslash}p{6em}|} 

			\hline
			Parameter Set & Amortized Processing Time ($\mu s$) & Amortized Stream Out Time ($\mu s$) & Software \cite{FHEWlike} ($\mu s$) & Improvement\\
			\hline
			MEDIUM	& 6615 & 49 & 141100 & 21.1$\times$\\
			\hline
			STD128\_AP & 13238 & 48 & 283800 & 21.3$\times$\\
			\hline
			STD192 & 26253 & 54 & 578400 & 22.0$\times$\\
			\hline
			STD256 & 52523 & 54 & 1180800 & 22.4$\times$\\
			\hline
			STD192Q & 52524 & 59 & 1270500 & 24.2$\times$\\
			\hline
		STD256Q & 70031 & 58 & 1571500 & 22.4$\times$\\
			\hline
		\end{tabular}
	\end{center}
	\label{tab:tb2}
\end{table}
\subsection{Measurement of Bootstrap of The Third Generation FHE}
\label{sec:measurementsub2}

The parameter sets used to benchmark our implementation of the third-generation FHE are listed in \autoref{tab:tb1} and adopted from \cite{FHEWlike}. Since our work only implements the homomorphic accumulation of the bootstrap process (including evaluation, accumulation, and key switch) on the hardware, we only report the measurement of this operation to emphasize our advancement. It is composed of two parts, the processing time and the time of streaming out the result to host. \autoref{tab:tb2} summarizes the measurement of the homomorphic accumulation function. Due to the pipelined nature of the proposed accelerator, the maximum parallelism is 12 accumulations. So. the reported time is amortized over 12 inputs. Because the homomorphic accumulation function is independent of the input binary gate, we do not differentiate it during the measurement. The reported time is averaged over all measured input gates.

The software implementation \cite{FHEWlike} of the FHEW scheme from the PALISADE library \cite{PALISADE} operates on the same host machine and is used for comparison. \autoref{tab:tb2} gives the comparison result of the proposed accelerator over software implementation. As stated above, only the homomorphic accumulation part is compared. On average, a $21\times$ speed-up for the homomorphic accumulation function is achieved compared with the software implementation.

\subsection{Measurement of The Proposed PSI}
\label{sec:measurementsub3}
In our implementation of the proposed PSI protocol, we set the encryption-related parameters to be $N=2048$, $log_{2}(Q)=54$, with $\sigma=3.19$, which achieves around 128-bit security level according to the LWE estimator \cite{LWEestimator}. The $B_G$ of the RGSW and $B_{KS}$ of the RLWE key-switch key are both set to $2^9$. Since the proposed PSI is not directly available in open-source libraries, we developed the necessary components of the scheme ourselves for baseline comparison. 

The average processing time of the two basic operations of the proposed PSI, RLWE substitution and $RLWE\bigotimes RGSW$, as deployed on the hardware are shown in \autoref{tab:tb4}. A comparison to our own software implementation is also included in the table. The raw measurement shows a speed-up factor of over 140, which is much higher compared to the improvement of the bootstrap process. A discussion of this discrepancy is incorporated in a later section. The last column, ‘Scaled Improvement,’ is added for this purpose and is discussed later, as well.
\begin{table}
	\caption{Comparison of The Processing Time of The Two Operations for The Proposed PSI}	
	\begin{center}
		\begin{tabular}{ |c|c|c|c|c| } 
			\hline
			& Proposed  & Software & & Scaled \\
			Operation & Accelerator ($\mu s$) & ($\mu s$) & Improvement & Improvement \\
			\hline
			RLWE Substitution	& 105 & 17616 & 167.8$\times$ & 27.9$\times$ \\
			\hline
			$RLWE\bigotimes RGSW$ & 105 & 14739 & 140.4$\times$ & 23.4$\times$ \\
			\hline
		\end{tabular}
	\end{center}
	\label{tab:tb4}
\end{table}
\begin{table}
	\caption{Sender's Processing Time and Communication Size of The Proposed PSI}	
	\begin{center}
		\begin{tabular}{ |c|c|c|c|c| } 
			\hline
			\multicolumn{2}{|c|}{} &  & \multicolumn{2}{|c|}{Communication Size}\\
			\multicolumn{2}{|c|}{Parameters} & Sender's Processing & \multicolumn{2}{|c|}{(MB)}\\
			\cline{1-2}\cline{4-5}
			b & k & Time ($s$) & b & k \\
			\hline
			\multirow{3}{*}{32} & 14 & 1642 & 27.0 & 256 \\
			\cline{2-5}
			& 12 & 814 & 7.5 & 64 \\
			\cline{2-5}
			& 10 & 585 & 2.1 & 16 \\
			\hline
			\multirow{3}{*}{30} & 14 & 1148 & 24.0 & 256 \\
			\cline{2-5}
			& 12 & 410 & 6.8 & 64 \\
			\cline{2-5}
			& 10 & 203 & 1.9 & 16 \\
			\hline
			\multirow{3}{*}{28} & 14 & 935 & 21.0 & 256 \\
			\cline{2-5}
			& 12 & 287 & 6.0 & 64 \\
			\cline{2-5}
			& 10 & 102 & 1.7 & 16 \\
			\hline
		\end{tabular}
	\end{center}
	\label{tab:tb5}
\end{table}

\begin{figure}
	\caption{Time Breakdown of the Proposed PSI.}
	\includegraphics[scale=0.5]{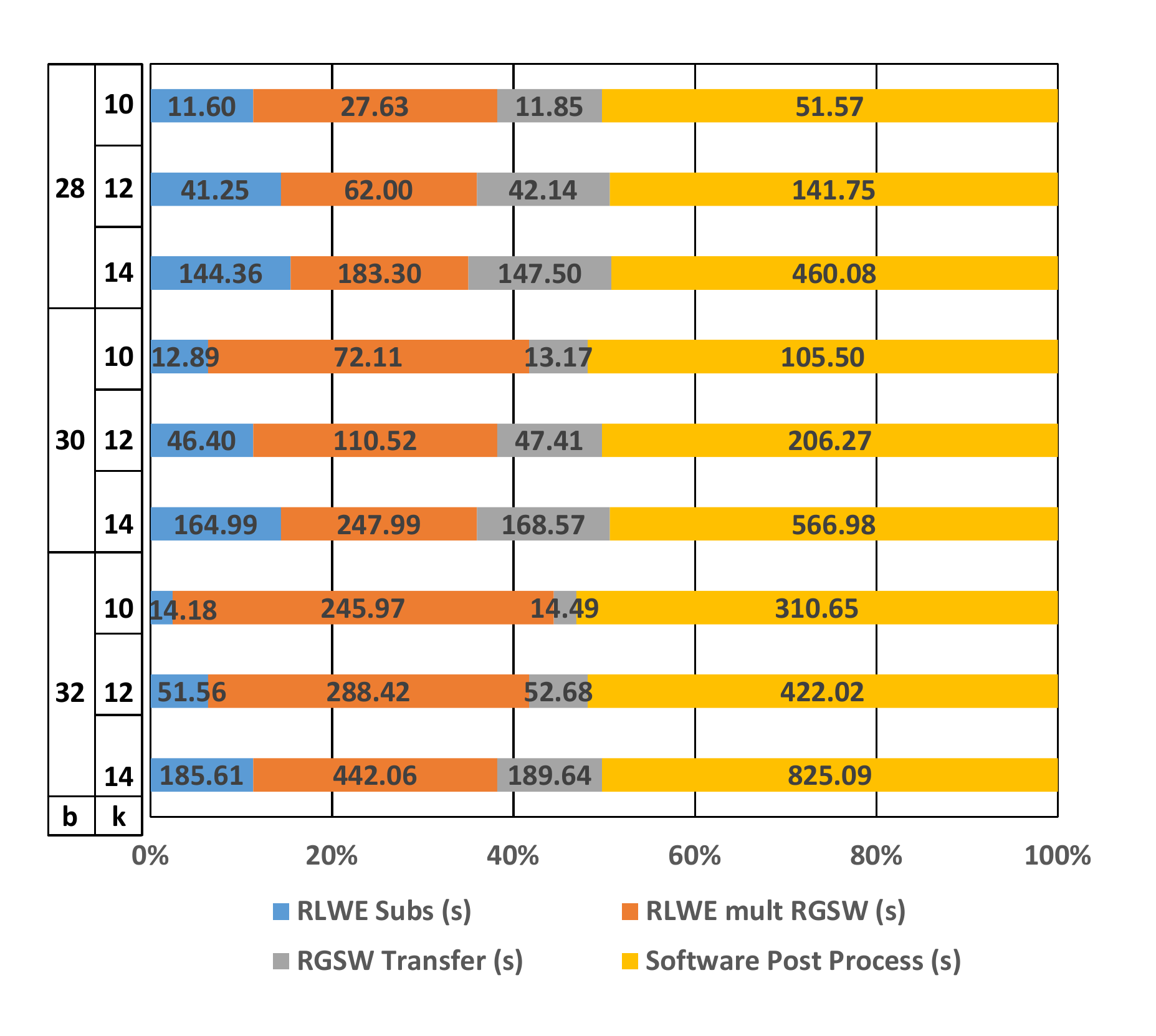}
	\centering
	\label{fig:fig19}
\end{figure}
Based on the time consumption of the basic operations from \autoref{tab:tb4}, the processing times on the Sender’s side with the proposed accelerator and the communication size of the proposed PSI are listed in \autoref{tab:tb5}. Since the complexity of our scheme is only directly dependent on the bit width $b$ and the hash table size $2^k$, assuming only one element in each bin on the Receiver’s side, we only list these two factors as design parameters in the table, with the security parameters set as above. In the Receiver-to-Sender communication size, the key-switch keys and the RGSW-encrypted $-\boldsymbol{z}$ are not included, which are of size 2.1 MB and 384 KB, respectively. Note that a modulus switch process can be applied to the returning LWE ciphertexts from the Sender to the Receiver, which can further reduce the message size by $15$\textasciitilde$20\%$ \cite{PSI6}. \autoref{fig:fig19} shows a time breakdown of the proposed PSI operating with the proposed accelerator. Four parts are included: (a) RLWE substitution; (b)$RLWE\bigotimes RGSW$; (c) RGSW transfer, which transfers the reconstructed RGSWs to the FPGA DDR; and (d) software post process. The first three are attributed to hardware. The measured time consumption of each part is also included in the diagram. The software post processing times are raw measurement data and not scaled, which will be discussed in next section.

\subsection{Analysis of The Measurement Results}
\label{sec:measurementsub4}
\subsubsection{Software Inefficiency Encountered during Measurement}
As mentioned in the above section, compared to the improvement of the bootstrap process listed in \autoref{tab:tb2}, we see a higher speed-up in the basic PSI operations, as shown in \autoref{tab:tb4}. The discrepancy mainly results from the different software implementations that are being used in the comparison. Since the proposed PSI and its basic operations are not directly available in open-source libraries, we developed the software implementation ourselves from scratch for both verifying the hardware design and baseline comparison. We also built our own software for the bootstrap process for the purpose of hardware verification and comparison. 

However, due to our relatively limited effort, our own software code may not perform as efficiently as the highly optimized open-source libraries. In order to estimate the potential software performance discrepancy, a comparison between our own software and an open-source library \cite{PALISADE} is conducted with the same host machine using commonly available operations such as NTT/INTT, polynomial operations, and bootstrap process. Based on the comparison, our own software code is around $6\times$ slower compared to open-source library. Hence, the measured improvement in the third column of \autoref{tab:tb4} is scaled by 6 to factor in potential software optimization for a more realistic speed-up number for the basic operations of the proposed PSI. This scaled number is shown in the last column of \autoref{tab:tb4}. 

The inefficiency in our software code includes unoptimized post processing, which takes about $50\%$ of the total processing time of the proposed PSI operating on the accelerator (\autoref{fig:fig19}). Thus, by factoring out this inefficiency, the total time consumption of the proposed PSI could be reduced by around $42\%$ (which is not accounted for in the reported performance in \autoref{tab:tb5}). 
\begin{table}
	\caption{Attainable Bound of Sender's Processing Time of The Proposed PSI}	
	\begin{center}
		\begin{tabular}{ |c|c|c|c| } 
			\hline
			\multicolumn{2}{|c|}{Parameters} & \multicolumn{2}{|c|}{Sender's Processing (s)}\\
			\hline
			b & k & Measured & Attainable Bound \\
			\hline
			\multirow{3}{*}{32} & 14 & 1642 & 273 \\
			\cline{2-4}
			& 12 & 814 & 135 \\
			\cline{2-4}
			& 10 & 585 & 97 \\
			\hline
			\multirow{3}{*}{30} & 14 & 1148 & 191 \\
			\cline{2-4}
			& 12 & 410 & 68 \\
			\cline{2-4}
			& 10 & 203 & 33 \\
			\hline
			\multirow{3}{*}{28} & 14 & 935 & 155 \\
			\cline{2-4}
			& 12 & 287 & 47 \\
			\cline{2-4}
			& 10 & 102 & 17 \\
			\hline
		\end{tabular}
	\end{center}
	\label{tab:tb6}
\end{table}
\subsubsection{I/O Bandwidth Bottleneck of the Implemented Accelerator}
During the measurement, we find that the latency of processing just one input on the proposed acceleration hardware is \textasciitilde350 $\mu s$ for RLWE substitution and 309 $\mu s$ for $RLWE\bigotimes RGSW$, which includes 120 $\mu s$ streaming in and out. Due to the pipelined nature of the proposed accelerator, a maximum parallelism of 13 can be achieved in the RLWE mode. Therefore, ideally, the average time consumption of processing one input on the hardware should be \textasciitilde17 $\mu s$, which is $6\times$ faster compared to the numbers listed in the first column of \autoref{tab:tb4}. This shows that, in the RLWE mode, the accelerator is bottlenecked by the I/O bandwidth. In the case that an optimized I/O is achieved, $6\times$ better performance can be extracted from the proposed accelerator. 

\autoref{tab:tb6} summarizes the (estimated) attainable bound of processing time of the proposed PSI, which both factors out software inefficiency and operates on an optimized I/O.

\section{Conclusion}
\label{sec:conclusion}
In conclusion, the first hardware acceleration architecture for third-generation FHE is proposed in this paper. Featuring an asymmetric INTT/NTT configuration, the proposed compute pipeline achieves less resource usage while maintaining a high throughput. An extensive analysis of the architecture is presented. An unbalanced PSI protocol based on third-generation FHE is also proposed to better demonstrate the architecture. Supplemented by several optimizations for reducing the communication and computation costs, the proposed PSI achieves a computation cost independent of the Sender’s set size. Implemented with AWS cloud FPGA, the proposed accelerator achieves over 21× performance improvement compared with a software implementation on various subroutines of the FHE and the proposed PSI at 125 MHz.

\section*{Acknowledgment}
Covered for blind review.

\appendix
\section{RLWE to LWE Conversion}
\label{sec:appRLWE2LWE}
Since RLWE is a special form of LWE, the coefficients of the polynomial $\boldsymbol{b}$ of an RLWE ciphertext $RLWE_{\boldsymbol{z}}(\boldsymbol{m})=[\boldsymbol{a,b}]$ can be converted into multiple separate LWE ciphertexts under the same secret key with some transformation of polynomial $\boldsymbol{a}$. For example, in an RLWE ciphertext $[\boldsymbol{a}=\sum\boldsymbol{a}[i] X^i,\boldsymbol{b}=\sum\boldsymbol{b}[i] X^i]$, the coefficient of $\boldsymbol{b}$ at index $0$:
\begin{equation}
	\boldsymbol{b}[0]=\boldsymbol{z}[0]\times\boldsymbol{a}[0]-\sum_{1}^{N-1}\boldsymbol{z}[i]\times\boldsymbol{a}[N-i]+\boldsymbol{e}[0]+\boldsymbol{m}[0],
	\label{eq:RLWE2LWE}
\end{equation}
can be viewed as an LWE ciphertext $LWE_z (\boldsymbol{m}[0])=[\boldsymbol{a_{LWE}},\boldsymbol{b}[0]]$ encrypted by secret key $\boldsymbol{z}$, where $\boldsymbol{a_{LWE}}=[\boldsymbol{a}[0],-\boldsymbol{a}[N-1],-\boldsymbol{a}[N-2],\cdots-\boldsymbol{a}[1]]$. 

\section{Example of RLWE Expansion with RLWE Substitution}
\label{sec:appRLWEexpan}
To demonstrate how RLWE substitution fulfills the expansion, $k=N+1$ is shown as an example. For $k=N+1$, $(X^i )^k=(-1)^i X^i$. Thus, the addition of the substituted ciphertext to the original one extracts the even index coefficients of the message $m$,  and the subtraction extracts the odd index coefficients, as shown in \autoref{eq:RLWEsubseven}.
\begin{equation}
	\begin{split}
		RLWE_{\boldsymbol{z}}(\sum\boldsymbol{m}[i]X^i )+RLWE_{\boldsymbol{z}}(\sum\boldsymbol{m}[i] (X^i)^k )&=RLWE_{\boldsymbol{z}}(\sum2\times\boldsymbol{m}[2i] X^{2i})\\
		RLWE_{\boldsymbol{z}}(\sum\boldsymbol{m}[i]X^i )-RLWE_{\boldsymbol{z}}(\sum\boldsymbol{m}[i] (X^i)^k )&=RLWE_{\boldsymbol{z}}(\sum2\times\boldsymbol{m}[2i+1] X^{2i+1})
		\label{eq:RLWEsubseven}
	\end{split}
\end{equation}

Therefore, by recursively substituting with $k=n/2^s+1$, for $s\in[0,\log_2N-1]$, each coefficient $\boldsymbol{m}[i]$ of the message is extracted into a separate RLWE ciphertext $RLWE_{\boldsymbol{z}} (N\times\boldsymbol{m}[i])$. The scale can be offset by pre-scaling the message with the multiplicative inverse of the $N$ in $Z_Q$.

\section{LUT Comparison with Permutation Based Hashing}
\label{sec:appPermhash}
The comparison in the homomorphic LUT still holds with permutation-based hashing. Assuming that $x_{iH}$ from the Receiver and $y_{jH}$ from the Sender are in the same bin after hashing and $x_{iH}=y_{jH}$, from \autoref{eq:permhash}, it is apparent that 
\begin{equation}
	\begin{split}
		pos(x_i)=&H(x_{iH})\ XOR\ x_{iL}\\
		=&H(y_{jH})\ XOR\ x_{iL}\\
		=&pos(y_j)\\
		=&H(y_{jH})\ XOR\ y_{jL}.\\
	\end{split}
	\label{eq:permhashdeduce}
\end{equation}
Therefore, $x_{iL}=y_{jL}$, resulting in $x=y$. Thus, the correctness of the LUT based PSI holds with permutation-based hashing.

\section{INTT Algorithm}
\label{sec:appINTT}
\begin{algorithm}[H]
	\caption{Inverse NTT | $INTT(\boldsymbol{a_{NTT}})$}\label{alg:alg1}
	\KwInput{$\boldsymbol{a_{NTT}}\in\mathbb{Z}_Q^N$ in bit reverse order, $Q\equiv1 \mod 2N$; a vector of twiddle factors $TF\in\mathbb{Z}_Q^N$ storing the powers of $\psi_N^{-1}$ in bit reverse order.}
	\KwOutput{$\boldsymbol{a}\leftarrow INTT(\boldsymbol{a_{NTT}})$ in INTT domain with normal order.}
	$t \gets 1$\;
	\For{$(m\gets N;\ m > 1;\ m\gets m/2)$}{
		$j_1 \gets 0$\;
		$h \gets m/2$\;
		\For{$(i\gets0;\ i<h;\ i++)$}{
			$j_2\gets j_1+t-1$\;
			$S\gets TF[h+i]$\;	
			\For{$(j\gets j_1;\ j\leq j_2;\ j++)$}{
				\tcp{Gentleman-Sande Butterfly}
				$U\gets \boldsymbol{a_{NTT}}[j]$\;  
				$V\gets \boldsymbol{a_{NTT}}[j+t]$\;
				$\boldsymbol{a_{NTT}}[j]\gets U+V \mod Q$\;
				$\boldsymbol{a_{NTT}}[j+t]\gets (U-V)\times S \mod Q$\;
			}
			$j_1\gets j_1+2t$\;
		}
		$t\gets 2t$\;
	}
	\For{$(j\gets 0;\ j<N;\ j++)$}{
		$\boldsymbol{a_{NTT}}[j]\gets \boldsymbol{a_{NTT}}[j]\times N^{-1}  \mod Q$\;
	}
	$\boldsymbol{a}\gets \boldsymbol{a_{NTT}}$\;
\end{algorithm}
\newpage
\section{NTT Algorithm}
\label{sec:appNTT}
\begin{algorithm}[H]
	\caption{NTT | $NTT(\boldsymbol{a})$}\label{alg:alg2}
	\KwInput{$\boldsymbol{a}\in\mathbb{Z}_Q^N$, $Q\equiv1 \mod 2N$; a vector of twiddle factors $TF\in\mathbb{Z}_Q^N$ storing the powers of $\psi_N$ in bit reverse order.}
	\KwOutput{$\boldsymbol{a_{NTT}}\leftarrow NTT(\boldsymbol{a})$ in NTT domain with bit reverse order.}
	$t \gets N$\;
	\For{$(m\gets N;\ m < N;\ m\gets 2m)$}{
		$t\gets t/2$\;
		\For{$(i\gets0;\ i<m;\ i++)$}{
			$j_1\gets 2\cdot i\cdot t$\;
			$j_2 \gets j_1+t-1$\;
			$S\gets TF[m+i]$\;	
			\For{$(j\gets j_1;\ j\leq j_2;\ j++)$}{
				\tcp{Cooley-Tukey Butterfly}
				$U\gets \boldsymbol{a}[j]$\;  
				$V\gets \boldsymbol{a}[j+t]\times S$\;
				$\boldsymbol{a}[j]\gets U+V \mod Q$\;
				$\boldsymbol{a}[j+t]\gets (U-V) \mod Q$\;
			}
		}
	}
	$\boldsymbol{a_{NTT}}\gets \boldsymbol{a}$\;
\end{algorithm}

\bibliographystyle{alpha}
\bibliography{abbrev3,crypto,biblio}

\end{document}